\documentclass[a4paper,onecolumn,twoside,accepted=2025-01-14]{quantumarticle}

\pdfoutput=1
\usepackage[utf8]{inputenc}
\usepackage[english]{babel}
\usepackage[T1]{fontenc}
\usepackage{amsmath}
\usepackage{hyperref}
\usepackage{cleveref}

\usepackage{tikz}
\usepackage{lipsum}

\usepackage{graphicx}
\usepackage{subcaption}
\usepackage{amsmath,amsfonts,amssymb,amsthm,amsbsy,mathtools}
\usepackage[normalem]{ulem}
\usepackage{bbold}
\usepackage{epic}
\usepackage{eepic}
\usepackage{mathrsfs}
\usepackage[all]{xy}
\usepackage[T1]{fontenc}
\usepackage[utf8]{inputenc}
\usepackage{color}
\usepackage{centernot}
\usepackage{mathtools}
\usepackage{stmaryrd}
\usepackage{physics}
\usepackage{comment}
\usepackage{endnotes}
\usepackage[numbers,sort&compress]{natbib}

\usepackage{mathtools}

\newcommand{\GLM}{purified measurement }
\newcommand{\HSL}{twirled observable }
\newcommand{\GLMshort}{PM }
\newcommand{\HSLshort}{TO }
\newcommand{\HSLshortnospace}{TO}
\newcommand{\GLMshortnospace}{PM}

\crefname{equation}{eq.}{eq.}

\newcommand{\K}[1]{\hat{K}_{SA}(#1)}

\def\e{\epsilon}

\def\e{\mathrm{e}}
\def\i{\mathrm{i}}

\newmuskip\pFqmuskip   

\newcommand*\pFq[6][8]{%
  \begingroup 
  \pFqmuskip=#1mu\relax
  \mathcode`\,=\string"8000
  \begingroup\lccode`\~=`\,
  \lowercase{\endgroup\let~}\pFqcomma
  {}_{#2}F_{#3}{\left[\genfrac..{0pt}{}{#4}{#5};#6\right]}
  \endgroup
}
\newcommand{\pFqcomma}{\mskip\pFqmuskip}
\newcommand{\piGlm}{\hat \Pi_{\text{\GLMshort}}}

\newcommand{\tens}{\otimes}

\newcommand{\dif}[2]{\frac{\partial#1}{\partial#2}}
\newcommand{\hal}{\mathcal{H}}
\newcommand{\del}{\partial}

\newcommand{\tot}{\hat H}
\newcommand{\hs}{\ket{\Psi}}
\newcommand{\kuch}{\text{Kucha\v{r}'s }}
\newcommand{\kuchar}{\text{Kucha\v{r} }}

\newcommand{\ketbrat}[1]{\ketbra{\phi_{#1}}{\phi_{#1}}_C}
\newcommand{\brakett}[2]{\braket{\phi_{#1}}{\phi_{#2}}_C}
\newcommand{\kett}[1]{\ket{\phi_{#1}}_C}
\newcommand{\brat}[1]{\bra{\phi_{#1}}_C}
\DeclareMathOperator{\sinc}{sinc}

\begin{document} 
\title{Measurement events relative to temporal quantum reference frames}

\author{Ladina Hausmann}
\affiliation{Institute for Theoretical Physics, ETH Zurich, Switzerland}

\author{Alexander Schmidhuber}
\affiliation{Institute for Theoretical Physics, ETH Zurich, Switzerland}

\author{Esteban Castro-Ruiz}
\affiliation{Institute for Theoretical Physics, ETH Zurich, Switzerland}
\affiliation{Université Paris-Saclay, Inria, CNRS, LMF, 91190 Gif-sur-Yvette, France}
\affiliation{Institute for Quantum Optics and Quantum Information (IQOQI) Vienna, Austrian Academy of Sciences, Boltzmanngasse 3, A-1090 Vienna, Austria}

\begin{abstract} 
\noindent The Page-Wootters formalism is a proposal for reconciling the background-dependent, quantum-mechanical notion of time with the background independence of general relativity. However, the physical meaning of this framework remains debated. In this work, we compare two consistent approaches to the Page-Wootters formalism to clarify the operational meaning of evolution and measurements with respect to a temporal quantum reference frame. The so-called ``twirled observable'' approach implements measurements as operators that are invariant with respect to the Hamiltonian constraint. The ``purified measurement'' approach instead models measurements dynamically by modifying the constraint itself. While both approaches agree in the limit of ideal clocks, a natural generalization of the purified measurement approach to the case of non-ideal, finite-resource clocks yields a radically different picture. We discuss the physical origin of this discrepancy and argue that these approaches describe operationally distinct situations. Moreover, we show that, for non-ideal clocks, the purified measurement approach yields a time non-local evolution equation, which can lead to non-unitary evolution. Moreover, it implies a fundamental limitation to the operational definition of the temporal order of events. Nevertheless, unitarity and definite temporal order can be restored if we assume that time is discrete.
\end{abstract}
	
\maketitle

\section{Introduction}

In standard quantum mechanics, systems evolve via the Schrödinger equation relative to an external background time. In general relativity, on the other hand, the metric determines the reading of rods and clocks, irrespective of any external background. The contrast between these two notions of time calls for a conceptual framework in which quantum systems evolve in the absence of an external time parameter.

The Page-Wootters formalism proposes a solution to this problem in non-relativistic quantum mechanics. Motivated by the
Wheeler-De-Witt equation \cite{Wheeler_1978},
\begin{equation} \label{eq:wdw}
    \hat{H} \ket{\Psi} = 0, 
\end{equation}
the Page-Wotters formalism \cite{Page_1983,Wootters_1984} poses that time evolution emerges from the correlations between the system of interest and a temporal reference frame -- a clock. 
The total system, composed of the system of interest and the frame, satisfies the so-called Hamiltonian constraint, \cref{eq:wdw}.
Because $\ket{\Psi}$ does not depend on any external background time, the approach based on \cref{eq:wdw} is referred to as ``timeless'' quantum mechanics. 
For a review of the Page-Wootters formalism, see \cref{appendix:page_wooters}.

Although the Page-Wootters formalism reproduces time evolution in a timeless setting, the formalism has been criticized \cite{Kuchar_2011, Unruh_1989}. 
In particular, Kucha{\v{r}} noted that the post-measurement states of the formalism do not satisfy \cref{eq:wdw} and predict inconsistent probabilities for two-time measurements \cite{Kuchar_2011}. 
A more detailed review of Kucha{\v{r}}'s criticisms is given in \cref{appendix:kuchar}.

In the current literature, there are two different ways to resolve Kucha{\v{r}}'s criticisms. 
The first route, which we call the \HSL (\HSLshortnospace) approach, has been taken by \cite{Hoehn_2021} (see also \cite{Dolby_2004}). 
It is based on time-translation invariant operators, which are inspired by Dirac quantization \cite{Dirac_2001,Henneaux_1992} and time-reparametrization invariant observables in quantum gravity \cite{Bojowald_2011,Bojowald_2011a,Gambini_2004,Gambini_2009, Rovelli_1990,Rovelli_2014,RovelliC_2009}. 
The second approach, which we call the \GLM (\GLMshortnospace) approach, was pursued by \cite{Giovannetti_2015}.
In this approach, the idea is to modify the constraint to describe the measurement dynamically (the same idea was also pursued in \cite{Hellmann_2007,Mondragon_2007,Aharanov1961}). 
More specifically, the modified constraint couples an ancillary system to the clock and the system to be measured, and a measurement on the ancilla reveals the outcome of the measurement on the system. 

Both approaches produce a theory that addresses \kuchar's criticisms. 
This leads to the question: are these two proposals physically equivalent? 
In this work, we compare both approaches and shed light on the operational meaning of each. 

Our analysis is motivated by recent literature on quantum reference frames transformations \cite{Castro_Ruiz_2020,Giacomini_2019,H_hn_2012,H_hn_2020, H_hn_2021, Giacomini_2021, Loveridge_2019, Hoehn_2019}. 
In these works, an important goal is to understand, intuitively speaking, how physics ``looks like'' as described from the perspective of a quantum reference frame. 
We explore this question in the case of temporal reference frames, and in particular, in the case where the clocks have finite resources, that is, when clocks are not ideal.

After presenting the clock models used in this work (\cref{sec:clocks}), we review the \HSL and \GLM approaches in \cref{sec:review}. 
We show that in physical scenarios where the clocks can be treated as ideal, both models give the same predictions, and are equivalent to ordinary quantum mechanics with an external background time.

Usually, clocks have a large amount of resources at their disposal, so the ideal clock approximation holds in most practical situations.
However, there are fundamental limitations to the ideal clock approximation, most notably from the perspective of gravity. 
Specifically, ideal clocks require an infinite energy spread, creating a large back reaction on spacetime. 
This leads not only to the well-known, fundamental operational limitations on spacetime measurements at the Plank scale \cite{Bronstein_2011}, but also to limitations stemming from gravitationally-induced entanglement between clocks \cite{Castro_Ruiz_2017}. 
Motivated by these considerations, we construct a natural generalization of measurements in the \GLMshort approach for non-ideal clocks in \ref{nonidealhslglm}. 
We obtain a time-non-local equation for the evolution of the system with respect to the clock, as reported in \cite{Smith_Ahmadi_2019}. 
We compare our model with its \HSLshort counterpart \cite{Hoehn_2021, de_la_Hamette_2021}, and show that the \HSLshort and \GLMshort approaches lead to different physical predictions. 

We argue that this difference arises because each approach has a different notion of measurement event\footnote{Throughout the paper, we use the term measurement event, or simply event, to refer to the notion of measuring a system at a certain time.}, and therefore describes a different experimental situation. 
Moreover, because ideal clocks are nothing but a limiting case of non-ideal clocks, we argue that the \HSLshort and \GLMshort approaches should be considered operationally different even in the ideal-clock case. 

In the \HSLshort approach, measurements and state-update proceed basically as in ordinary quantum mechanics with external time evolution. 
Importantly, time evolution with respect to the non-ideal clock is unitary, and the temporal order between events can be defined sharply. 
The extension of the \GLMshort model to non-ideal clocks leads to a radically different picture. 
Here, the clock can no longer resolve single time instances, leading to important measurable consequences. 
In particular, the time evolution of the system ``as seen'' by the clock is no longer unitary, and the probability rule is modified.
Moreover, we show that measurement events, understood as an interaction between the system of interest and an ancillary device, can occur in an indefinite temporal order, pointing to a fundamental limitation to the notion of the temporal order of events, with potential implications to the notion of indefinite causal structure \cite{Oreshkov_2012, Baumann_2022, Oreshkov_2019}. 

Finally, in \ref{sec:discrete}, we construct a discrete-time version of the \GLMshort model and show that unitarity is recovered for general measurement interactions. 
If unitary evolution is to be held at all costs, our analysis can be seen as an argument in favour of the idea that time is fundamentally discrete. 

\section{Ideal and non-ideal clocks}\label{sec:clocks}
Before analysing the \HSLshort and \GLMshort approaches, we summarize the clock models we use in this paper. Our summary is based on Ref. \cite{Hoehn_2021}.

In this work, a clock is a quantum system $C$ with Hilbert space $\mathcal H_C$\footnote{Throughout the paper we assume that the Hilbert space of the clock is given by $L^{2}[E,-E]$.} and a Hamiltonian 
\begin{equation} \label{ClockHamiltonian}
\hat{H}_C  = \int_{\sigma} \mathrm{d}e \, e \ketbra{e}{e}_C,
\end{equation}
where $\sigma = [-E, E]$ is the spectrum of the clock, for some energy $E$\footnote{
In the main text, we use a symmetric spectrum as the overlap $\brakett{t}{t'}$ is the widely-known sinc function, thus making formulas where $\brakett{t}{t'}$ appears more intuitive. The derivations of the results of sections \cref{nonidealhslglm} are valid for a general spectrum, except the explicit solution calculated in \cref{eq:sol_delta}, which can easily be generalized to general spectra. In \cref{sec:discrete}, we use the symmetric spectrum for constructing a clock with a bounded spectrum whose time states are orthogonal at the desired values.}. 
The measurement of time corresponds to a covariant positive operator-valued measure (POVM). 
In the case of time, a covariant POVM consists of a set of positive operators $\hat{\phi}_t$, for $t \in \mathbb{R}$, satisfying $e^{-i t \hat{H}_C }\hat{\phi}_{t^\prime} e^{i t \hat{H}_C } = \hat{\phi}_{t + t^\prime}$ and forming a resolution of the identity. 
In this work, we take $\hat{\phi}_t = \ketbrat{t}$, with
\begin{equation}
    \kett{t} = \frac{1}{\sqrt{2 E}}\int_{-E}^{E} e^{-i e t} \ket{e}_C.
\end{equation}
The states $\kett{t}$ transform as $e^{-i t^\prime \hat{H}_C }\kett{t} = \kett{t^\prime + t}$, and resolve the identity as
\begin{equation}
N_C \int dt  \ketbrat{t} = \mathbb{1}_{C}, 
\end{equation}
for a normalization constant $N_C = E/\pi$. 

The overlap between two clock states at different times $\brakett{t}{t'}$ is
\begin{equation}\label{overlap}
   N_C \brakett{t}{t'} = \frac{\sin(E(t-t'))}{\pi (t-t')}.
\end{equation}
It is easy to see from \cref{overlap} that the distinguishability of clock states for two different times, $\kett{t}$ and $\kett{t^\prime}$, increases as the energy $E$ increases. 
In this sense, energy is a resource for temporal quantum reference frames. 
A clock with finite $E$ is then an instance of a quantum reference frame with finite resources \cite{Ahmadi_2013}.
In the limit $E \rightarrow \infty$, the spectrum of the clock becomes $\sigma_C = \mathbb R$. 
In this special case, we denote the clock states by $\ket{t}_C$. In contrast to the case of finite $E$, the states $\ket{t}_C$ can be perfectly distinguished, and we can write their inner product as
\begin{equation}\label{idealip}
\braket{t}{t'}_C = \delta(t-t'). 
\end{equation}
In this work, we refer to a clock defined by the states $\ket{t}$ as \emph{perfect} or \emph{ideal}. 
For ideal clocks, the Hamiltonian in \cref{ClockHamiltonian} has a Hermitian conjugate \emph{time operator} $\hat{T}_C$, satisfying $[\hat T_C,\hat H_C] = i$, and $\hat T_C \ket{t}_C = t \ket{t}_C$. 
Naturally, ideal clocks do not exist. 
Whenever we say a clock is ideal, what we really mean is that we are considering an experiment in a regime where the approximation of \cref{idealip} is physically meaningful. 
On the other hand, we refer to the (more realistic) clocks corresponding to states $\kett{t}$ as \emph{imperfect} or \emph{non-ideal}.

\section{Review of the \HSL and \GLM approaches}\label{sec:review}
We briefly summarize the \HSLshort and \GLMshort approaches for the case of ideal clocks, as defined in \cref{sec:clocks}. 
We show that, in this idealized setting, both approaches lead to the same physical predictions. 

\subsection{Twirled observable approach}\label{sec:HSL}
\noindent Let us start with the \HSLshort approach, introduced in \cite{Hoehn_2021}. For a more detailed account, see \cref{appendix:perfect_clocks}.
The key point of this approach is to substitute the standard Schrödinger equation with a constraint 
\begin{equation}\label{PerfectConstraint}
(\hat{H}_C + \hat{H}_S )\ket{\Psi} = 0.
\end{equation} 
The dynamics of the system $S$, with Hamiltonian $\hat H_S$, is described by its correlations with a clock $C$. 
The Hamiltonian of the clock, $\hat{H}_C$, is ideal in the sense of  \cref{sec:clocks}.

The constraint recovers the Schrödinger equation by contracting \cref{PerfectConstraint} with $\bra{t}$. 
Solving the resulting equation for the initial condition $\ket{\psi_0}_S$ at $t=0$, we find the solution to \cref{PerfectConstraint}:
\begin{equation}\label{hslhistorystate}
    \ket{\Psi} = \int_{\mathbb R} \mathrm{d}t \, \ket{t}_C \otimes e^{-it \hat{H}_S} \ket{\psi_0}_S.
\end{equation}
Note that the evolution of $S$ with respect to $C$ is unitary. 
By this, we mean that, for any pair of times $t_1$ and $t_2$, the state of $S$ at $t_1$ is transformed into the state of $S$ at $t_2$, by a unitary operator. 
That is, $\ket{\psi(t_2)}_{S} = U(t_2, t_1) \ket{\psi(t_1)}_{S}$. Moreover, we require that the operators $U(t, t')$ satisfy the so-called group property: $U(t_3, t_2)U(t_2, t_1) = U(t_3, t_1)$.
Measurable physical quantities correspond to operators that commute with the constraint $\hat{H}_S + \hat{H}_C$, as they should leave the state in the space of states respecting the constraint. 
For any operator $\hat F_S$ on $S$ and any time $\tau$ we define the relational Dirac operator
\begin{equation}
\label{eq:HSLop}
\hat F_{CS}(\tau) =  \int_{\mathbb R} dt \, \ e^{- i t\hat H_C} \, \ketbra{\tau}{\tau}_C \, e^{ i t\hat H_C}\tens e^{- i t\hat H_S}\, \hat F_S \, e^{ i t\hat H_S}.
\end{equation} 
Note that $\hat F_{CS}(\tau)$ relies on the well-known $\mathcal{G}$-twirl procedure (see, e.g. \cite{Bartlett_2007}) for constructing invariant observables -- hence the name ``\HSL (\HSLshortnospace)'' approach.
By construction, $\hat F_{CS}(\tau)$ commutes with the constraint, so the state $\hat F_{CS}(\tau) \ket{\Psi} $ satisfies \cref{PerfectConstraint} as well. 

We can then define probabilities for consecutive measurements analogously to ordinary quantum mechanics. 
For example, let $\hat \Pi^k_S$ and $\hat \Pi^q_S$ be projectors corresponding to measurement outcomes $k$ and $q$, respectively, of two observables $\hat K_S$ and $\hat Q_S$. 
Assume that we have obtained the outcomes $k$ and $q$ at times $\tau_1$ and $\tau_2$, respectively. 
Then, in this approach, the operators $\ketbra{\tau_1}{\tau_1}_C \otimes \hat \Pi^k_S$ and $\ketbra{\tau_2}{\tau_2}_S \otimes \hat \Pi^q_S$ cannot be directly applied to the state as they are not relational, and would lead to the problems \kuchar pointed out.
Instead, we should use the corresponding relational Dirac operators from \cref{eq:HSLop}, $\hat \Pi^k_{CS}(\tau_1)$ and $\hat \Pi^q_{CS}(\tau_2)$, labelled by times $\tau_1$ and $\tau_2$.
The joint probability of measuring $p$ followed by $q$, at times $\tau_1$ and $\tau_2$ respectively, is given by 
\begin{align}\label{HSLProbRule}
P(k,q \vert \tau_1 \tau_2) = \bra{t=0}_C\bra{\psi_0}_S \hat \Pi^k_{CS}(\tau_1) \hat \Pi^q_{CS}(\tau_2) \hat \Pi^k_{CS}(\tau_1) \ket{\Psi}.
\end{align}
A straightforward calculation yields 
\begin{align}\label{qmprob}
    P(k,q \vert \tau_1 \tau_2) = \big \| \hat \Pi^q_S \mathrm{exp}(-i (\tau_2-\tau_1) \hat{H}_S) \hat \Pi^k_S \mathrm{exp}(-i \tau_1 \hat{H}_S) \ket{\psi_0}\big \|^2.
\end{align}
Therefore, if $\tau_2 > \tau_1$, \cref{HSLProbRule} gives the standard quantum mechanical probabilities for measuring $k$ at time $\tau_1$ and then $q$ at time $\tau_2$.

\subsection{Purified measurement approach}\label{sec:perfect_glm}
Instead of constructing covariant observables, the \GLMshort approach introduced in \cite{Giovannetti_2015} describes measurements dynamically, at the level of the constraint. In this sense, we say that the measurement is ``purified'' -- hence the name ``\GLM (\GLMshortnospace)'' approach. Specifically, the measurement is modelled by adding to the constraint an interaction term between the clock, the system and an ancilla: 
\begin{align}\label{PerfectGLMHamiltonian}
   \hat H &\ket{\Psi}_{CSA} = \left(\hat H_C + \hat H_S + \int d t \ketbra{t}{t}_C \otimes \K{t} \right)\ket{\Psi}_{CSA} = 0.
\end{align}
The ancillary system $A$\footnote{The Hilbert space of the system $A$ is the span of all possible measurement outcomes $\ket{r}_A$ that are all mutually orthogonal.} can be thought of as the pointer of a measurement device that records the measurement outcome. $\hat H_C$ is the Hamiltonian of a perfect clock. 

A standard choice for the measurement interaction is $\K{t} = \delta(t-\tau) \hat H_{SA}$, which models a measurement interaction happening sharply at time $t=\tau$. 
If we choose $\hat H_{SA}$ such that $e^{-i \hat H_{SA}} \ket{\psi(\tau)}_S \otimes \ket{r}_A \mapsto \sum_{a}\hat K^{a}_S\ket{\psi(\tau)}_S \otimes \ket{a}_A$, then this interaction is a purification of the POVM $\{K^{a}_S\}_a$.
We assume this interaction for the rest of this section. 
The states satisfying the constraint $\hat H \ket{\Psi} = 0$ and boundary conditions such that $\bra{t}_C\ket{\Psi}_{CSA} = \ket{\psi(t)}_S \otimes \ket{r}_A$, for $t \leq \tau$, have the form
\begin{align}
    \begin{split}
	\ket\Psi_{CSA} &= \int_{-\infty}^{\tau} dt \ket{t}_C \tens \ket{\psi(t)}_S \tens \ket{r}_A \\ & \quad + \int_{\tau}^{\infty} dt  \ket{t}_C \tens \hat U_S(t-\tau) e^{-i \hat H_{SA}}\ket{\psi(\tau)}_S \tens \ket{r}_A \\
    &= \int_{-\infty}^{\tau} dt \ket{t}_C \tens \ket{\psi(t)}_S \tens \ket{r}_A \\& \quad + \int_{\tau}^{\infty} dt  \ket{t}_C \tens \hat U_S(t-\tau) \sum_{a}\hat K^{a}_S\ket{\psi(\tau)}_S \otimes \ket{a}_A,
	\label{1meshs}
 \end{split}
\end{align}
which can be interpreted as follows. The system first evolves freely according to $\hat H_S$. 
The ancilla is assumed to start in a ``ready'' state $\ket{r}_A$ and evolve trivially until the measurement interaction, which occurs when the clock shows time $\tau$, resulting in the application of the unitary $\mathrm{exp}(-i \hat H_{SA})$. 
Afterwards, the system evolves freely again. 
By adding more measurement interactions with additional memories to the Hamiltonian we can model multi-time measurements. 

The probability that the measurement outcome $a$ has been written into the memory of the measurement apparatus at time $t$ is given by
\begin{align}
	\label{1prob}
	P(a|\tau)_t = \frac{\bra\Psi \ketbra{t}{t}_C \otimes \mathbb{1}_S \otimes \ketbra{a}{a}_A  \hs}{\bra\Psi \ketbra{t}{t}_C \otimes \mathbb{1}_S \otimes \mathbb{1}_A\hs}.
\end{align}
For ideal clocks, the denominator of \cref{1prob} is independent of $t$. As we will see in Section~\ref{nonidealhslglm}, this is no longer the case when the perfect-clock approximation is not valid.
(For alternative probability rules applied to Wigner's Friend scenarios within the Page and Wootters formalism, see \cite{baumann2021generalized,Trassinelli2022}.) 
The rule of \cref{1prob} can be naturally extended to multiple-time measurements by adding one ancillary system for each measurement \cite{Giovannetti_2015}. 
If we construct a purification for the observables $K_S$ and $Q_S$, set $t > \tau_2 > \tau_1$, it is easy to see that \cref{1prob} leads to \cref{qmprob}, showing that both frameworks give the same predictions for ideal clocks. 
A constructive proof of this result, showing how the two approaches are formally related, can be found in \cref{appendix:glm=hls}.

Note that the time $t$ is not the time of the modelled measurement interaction, but the time at which the register is measured. 
This only needs to happen once, after all of the measurement interactions have occurred.

From the joint probability distribution for all outcomes, we can construct the conditional probabilities $P((a|\tau)|(b|\tau'))_t$, which is the probability of measuring $a$ at time $\tau$, given that we measured $b$ at time $\tau'$. 
If we omit the subscript $t$ in $P((a|\tau)|(b|\tau'))_t$, then mean that we evaluate $P((a|\tau)|(b|\tau'))_t$ at a time $t \gg \tau, \tau' $. 
Both the \HSLshort and the \GLMshort approaches yield the same conditional probabilities in the case of perfect clocks. 
More precisely, consider two subsequent measurements on a system and let $P_I((b|\tau')|(a|\tau))$ denote the probability of observing the measurement result $\ket{b}$ at time $\tau'$ conditioned on measuring outcome $\ket{a}$ at time $\tau$. 
Here, $I \in \{\text{\GLMshort}, \text{\HSLshort}\}$ is an index determining whether the probability was computed in the \GLMshort or \HSLshort approach. 
Then, as we show in \cref{appendix:glm=hls}, $P_{\GLMshort}((b|\tau')|(a|\tau)) = P_{\HSLshort}((b|\tau')|(a|\tau))$. However, as our analysis of the non-ideal case below will show, that the approaches are operationally different.

\section{Non-ideal clocks}\label{nonidealhslglm}

In this section, we study the case where the idealization of a perfect clock is no longer valid. How do we describe the dynamics of a system with respect to a finite-resource temporal reference frame? To answer this question, we focus on the clock model defined by \cref{ClockHamiltonian} in the case where the energy $E$ is finite.
We show that the answer depends on whether we use the \HSLshort or the \GLMshort approach, as in this finite case the two approaches give different physical predictions. 
The non-ideal \HSLshort approach was studied in \cite{Hoehn_2021}; the non-ideal \GLMshort approach is developed in \cref{nonidealGLM}. 

\subsection{Twirled observable approach for non-ideal clocks}
The central aspects of the \HSLshort approach do not depend on whether the clocks are ideal or not. 
In particular, the constraint is still given by \cref{PerfectConstraint}, but with the bounded-spectrum Hamiltonian $\hat{H}_C$ replacing the ideal one. 
Likewise, covariant extensions of operators are defined by \cref{eq:HSLop}, and the probability rule remains the one of \cref{HSLProbRule} replacing the ideal clock with a non-ideal one.
This replacement results in a restriction of the system's energy values to $\sigma(\hat H_{S}) \cap (-\sigma)$, where $\hat H_S$ is the Hamiltonian of the system appearing in the constraint. 
Consequently, also the evolution of the system with respect to the clock is generated by a Hamiltonian which has a spectrum given by $\sigma(\hat H_{S}) \cap (-\sigma)$.
This restriction is derived in detail in Section IV A of \cite{Hoehn_2021}. 
Broadly, it is an example of the system being constrained to the physical system Hilbert space as a result of a non-ideal quantum reference frame in the perspective-neutral approach \cite{de_la_Hamette_2021}. 
As in the ideal case, the evolution of the system with respect to the clock is unitary.

\subsection{Purified measurement approach for non-ideal clocks}\label{nonidealGLM} 

In contrast to the \HSLshort approach, the \GLMshort approach is fundamentally different when clocks are non-ideal. Even with a natural extension of \cref{PerfectGLMHamiltonian}, as the one presented below, the evolution equation changes substantially, and as a result the evolution can become non-unitary. 
The probability rule remains formally the same as \cref{1prob} (changing $\ket{t}_C \rightarrow \kett{t}$), with the important difference that the denominator is now not always equal to 1. 

\subsubsection{The Hamiltonian constraint}
A natural generalization of \cref{PerfectGLMHamiltonian}, for the case of non-ideal clocks reads 
\begin{widetext}
\begin{equation}\label{eq:const_imperfect_glm}
    \hat H \ket{\Phi}_{CSA} = \left(\hat H_C + \hat H_S + N_{C} \int dt \ketbrat{t} \otimes \K{t} \right) \ket{\Phi}_{CSA} = 0,
\end{equation}
\end{widetext}
where $N_C \ketbrat{t}$ is an element of the clock's covariant POVM described in \cref{sec:clocks}, which corresponds to the measurement of time. 
We can then interpret the last term of \cref{eq:const_imperfect_glm} as a simplified description that an external observer would give of an internal observer looking at her clock and triggering a measurement interaction $\K{t}$, which depends on the time $t$ that she observes.

\subsubsection{Evolution equation}
As in the case of perfect clocks, we can also consider the state $\ket{\Psi}_{CSA}$ conditioned on the clock showing the time $t$, $\ket{\psi(t)}_{SA} = \brat{t}\ket{\Psi}_{CSA}$. 
Similar to the ideal case, we show in \cref{appendix:calc_imperfect} that $\ket{\Psi}_{CSA} = N_C \int \mathrm{d}t \kett{t} \otimes \ket{\psi(t)}_{SA}$ fulfils the constraint of \cref{eq:const_imperfect_glm} if and only if the conditioned state $\ket{\psi(t)}_{SA} = \brat{t} \ket{\Psi}_{CSA}$ satisfies the equation
\begin{align}\label{eq:imperfect_evolution}
\begin{split}
    i \frac{d}{dt} \ket{\psi(t)}_{SA} &= \hat H_S \ket{\psi(t)}_{SA} + N_C\int dt' \brakett{t}{t'}\K{t'}\ket{\psi(t')}_{SA}.
\end{split}
\end{align}
If the clock is non-ideal, then \cref{eq:imperfect_evolution} is non-local in time. By this, we mean that $\frac{d}{dt} \ket{\psi(t)}_{SA}$ does not depend only on the state of $SA$ at time $t$, but also on the state evaluated at different times.\footnote{Note that a time non-local \emph{equation} does not necessarily mean a time non-local or a non-unitary \emph{evolution}. Indeed, as shown in \cite{Smith_Ahmadi_2019}, in some cases a time non-local equation can be recast in the form of an ordinary Schr\"odinger equation, which is local in time and generates unitary time evolution in the sense of \cref{sec:review}, but features a different Hamiltonian $\hat h(t)$. However, to find the Hamiltonian $\hat h(t)$, in general, one needs to solve the time non-local equation first. }

We now turn to the solution of \cref{eq:imperfect_evolution} and show that, for non-ideal clocks, the \HSLshort and \GLMshort approaches are physically inequivalent. For simplicity, we consider the case where $H_S = 0$, as this case already captures the conceptual issues relevant to this work. For the general case, we can simply substitute $\K{t}$ with $\K{t} + \hat H_S$.

\begin{figure}
    \centering
    \includegraphics[scale=0.8]{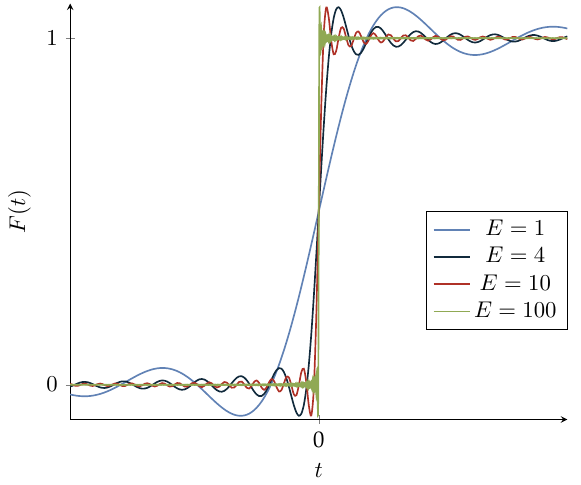}
    \caption{The function $F(t)$ for a clock with spectrum $\sigma_C = [-E,E]$, for different values of E. $F(t)$ parametrizes the solutions of \cref{eq:imperfect_evolution} in terms of the spectrum of the clock and characterizes 
    the time-non-locality of the evolution equation relative to $C$.}
    \label{fig:approx_theta}
\end{figure}

We solve \cref{eq:imperfect_evolution} perturbatively in \cref{appendix:calc_imperfect} for the case of periodic and non-periodic clocks, using similar methods to those of \cite{Smith_Ahmadi_2019}. 
We find
\begin{widetext}
\begin{equation}\label{eq:sol_glm_imperfect}
    \ket{\psi(t)} = \sum_{N = 0}^{\infty} (-i)^{N}\int_{\mathbb{R}}\dots \int_{\mathbb{R}} dt_N \dots dt_1 F(t-t_N) \dots F(t_2-t_1) \K{t_N} \dots \K{t_1} \ket{\psi_0},
\end{equation}
\end{widetext}
where $F(t) = N_C \int_{- \infty}^{t} dt' \bra{\phi_{t'}}\ket{\phi_0}$. In general, the evolution corresponding to \cref{eq:sol_glm_imperfect} is non-unitary.
Consider the case $\K{t} = \hat K_{SA} \delta(t-\tau)$. Then, we have
\begin{align}\label{eq:sol_delta}
    \ket{\psi(t)} &= \ket{\psi_0} + 2F(t-\tau) \sum_{N = 1}^{\infty} (-i)^{N}  \frac{1}{2^{N}} \hat K_{SA}^N \ket{\psi_0} \\&= \left(1 - F(t-\tau)\frac{i \hat K_{SA}}{1+\frac{i}{2} \hat K_{SA}}\right) \ket{\psi_0}.
\end{align}
The evolution $1 - F(t-\tau)\frac{i \hat K_{SA}}{1+\frac{i}{2} \hat K_{SA}}$ is only unitary for $F(t-\tau) = 0,1$, which is the case for $t \gg \tau$ or $t \ll \tau$. In other words, the evolution is unitary far away from the measurement. This already shows that the \HSLshort and \GLMshort approaches are physically inequivalent, as evolution in the \HSLshort approach is always unitary.

Moreover, $F(t)$ can be interpreted as a measure of the time-non-locality of the evolution equation (see \cref{eq:imperfect_evolution}) and its solution, \cref{eq:sol_glm_imperfect}. Taking the $E \longrightarrow \infty$ in \cref{eq:sol_glm_imperfect} enforces $F(t) = \Theta(t)$, where $\Theta(t)$ denotes the Heaviside theta function, and yields unitary evolution:
\begin{equation}
\ket{\psi(t)} = \overset{\leftarrow}{T} e^{-i \int_0^t \mathrm{d}s K_{SA}(s)} \ket{\psi_0}, 
\end{equation} 
where $\overset{\leftarrow}{T}$ denotes the time-ordering operator.
After taking the limit $E \longrightarrow \infty$, it is valid to take the limit of $\K t$ being an instantaneous interaction, i.e. 
$\K t = \delta(t-\tau) \hat H_{SA}$.\footnote{Importantly, the limit $E \longrightarrow \infty$ and $\K t \to \delta(t-\tau) \hat H_{SA}$ should be taken precisely in that order, as a clock with finite resources cannot control an operation at a sharp time $\tau$.} In this limit, we obtain the solution of \cref{sec:perfect_glm}, which gives the same predictions of the \HSLshort approach. 
However, away from this limit, $F(t)$ induces measurement events that are ``spread out'' in time. 
This is another aspect in which the \GLMshort approach departs from the \HSLshort approach in the non-ideal case, as measurement events are always ``sharp'' in the \HSLshort approach. 

In \cref{fig:approx_theta}, we plot $F(t)$ for different values of $E$ and illustrate how it approaches Heaviside's function. 
As $E$ increases, the spectrum broadens and the clock becomes more resourceful. 
As a consequence, time-non-local effects in the evolution equation are suppressed. 
Note, however, that $E$ cannot fundamentally increase indefinitely, due to the back-action of the clock on spacetime. 
While the problem is not apparent in this work, where gravity is not modelled, it is known that gravitational effects will eventually become relevant \cite{Bronstein_2011,Castro_Ruiz_2017}.
This is one of our motivations for considering non-ideal clocks. Therefore, the limit $E \longrightarrow \infty$ should be interpreted only as an approximation.
Roughly, this approximation is suitable when the minimum time $\delta t$ such that $\braket{\phi_0}{\phi_{\delta t}}\approx 0$ is much smaller that the minimum time $\Delta t$ such that $\braket{\psi(t)}{\psi(t + \Delta t)}_{SA} \approx 0$, if we assume that $\ket{\psi(t)}_{SA}$ evolves with the time-dependent Hamiltonian $\hat K_{AS} (t) + H_S$. 

\begin{figure}
    \centering
    \includegraphics[scale=0.8]{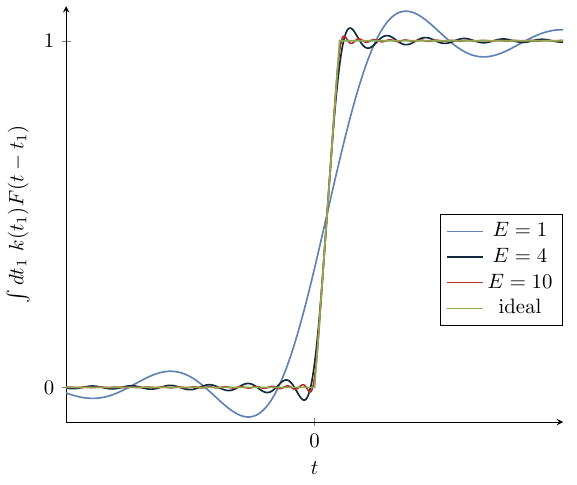}
    \caption{Magnitude of the order term of the time evolution $\int dt_1 \K{t_1} F(t-t_1)$ for $\K{t}= k(t) \hat{K}$, for clocks with spectrum $\sigma_C = [-E,E]$, with $k(t) = \chi_{[0,1]}(t)$, where $\chi$ is the characteristic function.  As expected, we see that for a broader spectrum of the clock, the non-ideal solutions approximate the ideal case better.}
    \label{fig:first_order}
\end{figure}

\begin{figure}
\centering
\begin{subfigure}{0.4\textwidth}
    \includegraphics[width=\textwidth]{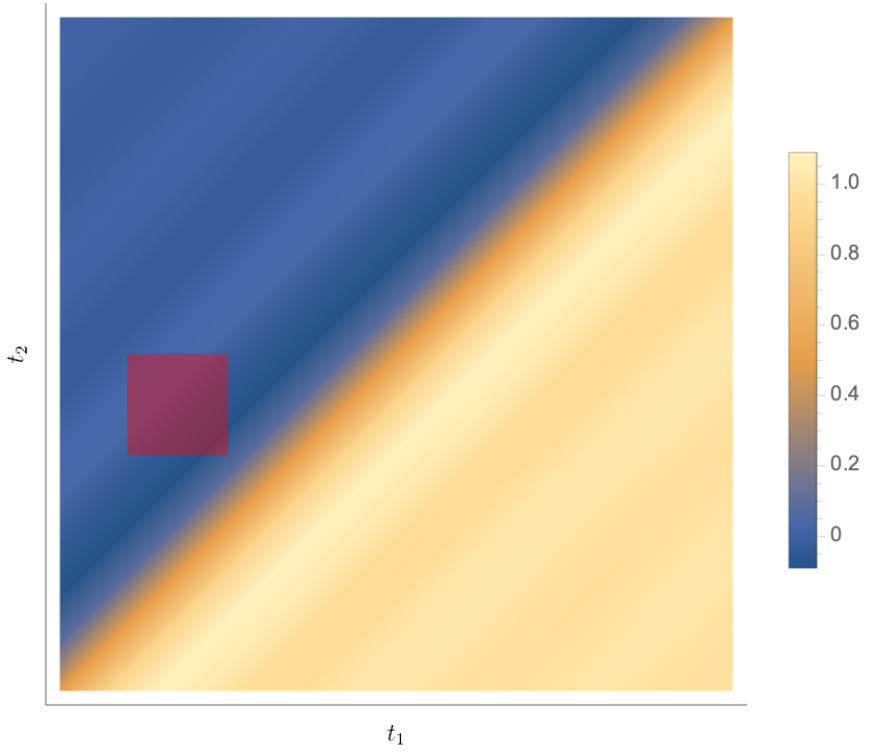}
    \caption{Non-ideal clocks}
    \label{fig:first}
\end{subfigure}
\hfill
\begin{subfigure}{0.4\textwidth}
    \includegraphics[width=\textwidth]{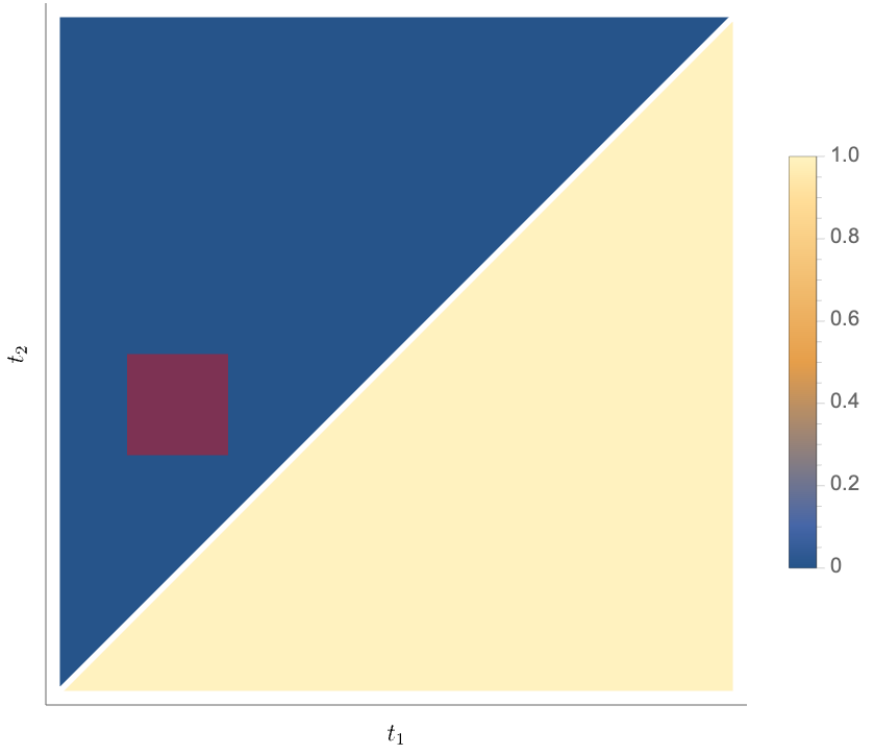}
    \caption{Ideal clocks}
    \label{fig:second}
\end{subfigure}
\hfill        
\caption{
The heat map shows the expression $F(t-t_1) F(t_1-t_2)$ occurring in \cref{eq:sol_expansion} in the case of non-ideal (left) and ideal (right) clocks. 
The red square illustrates a region where the contribution of the term $k(\tau_1-{t_1}) \hat K_1 k(\tau_2-{t_2})\hat K_2$ of \cref{eq:sol_expansion}, with $k(t) = \chi_{[0,1]}(t)$, is not zero. 
Here, the two times $\tau_1$ and $\tau_2$ are far apart enough to ensure that the support of $k(t-\tau_1)$ is disjoint from the support of $k(t-\tau_2)$. 
As these supports do not overlap in time, a non-zero contribution of $k(\tau_1-{t_1}) \hat K_1 k(\tau_2-{t_2})\hat K_2$ means an indefinite temporal order of events. 
The integration over $t_1,t_2$ of $F(t-t_1) F(t_1-t_2)k(\tau_1-{t_1}) k(\tau_2-{t_2}) \hat K_1 \hat K_2$ gives the contribution of this term. 
In the ideal clock case, we can see that, because there is no overlap in time of the two measurements, this term does not contribute. 
In the non-ideal case, the situation is different. Because the clock states are not orthogonal, even if there is no overlap of the two measurements in time, the contribution $k(\tau_1-{t_1}) \hat K_1 k(\tau_2-{t_2})\hat K_2$ to \cref{eq:sol_expansion} is not zero.} 

\label{fig:two_measurements}
\end{figure}

\subsubsection{Indefinite temporal order of events} \label{indeftemp}
The bounds on the energy of our clock have important consequences on the time ordering of events, understood as the measurement interactions triggered by the clock. 
To see this, consider two measurement interactions, modelled by $\K{t} = k(t-\tau_1) \hat K_1 + k(t-\tau_2) \hat K_2$ in \cref{eq:const_imperfect_glm}. 
For the following discussion, we assume that $\tau_1 < \tau_2$.
We now look at the state of the system ``as seen'' by the clock in this scenario. 
To second order in the expansion of \cref{eq:sol_glm_imperfect}, we have
\begin{align}\label{eq:sol_expansion}
\begin{split}
     \ket{\psi(t)} &= \ket{\psi_0} + \Big[ 
    \int dt_1 F(t-t_1) \left(k(t_1-{\tau_1}) \hat K_1 + k(t_1-{\tau_2}) \hat K_2\right) \\
    &+ \int dt_1 dt_2 F(t-t_1) F(t_1-t_2)\big(k(t_1-{\tau_1}) \hat K_1 
    \\&+ k(t_1-{\tau_2}\big) \hat K_2 \big) \left(k(t_2-{\tau_1}) \hat K_1 + k(t_2-{\tau_2}) \hat K_2 \right) 
    \Big] \ket{\psi_0}
\end{split}
\end{align}
In \cref{fig:first_order} we plot the magnitude of the first order term for one of the measurements in the case where $k(t) = \chi_{[0,1]}(t)$.
To second order, this expansion contains terms where $\hat K_1$ is applied before $\hat K_2$, that is, terms of the form
\begin{equation}\label{acausal}
    \int dt_1 dt_2 F(t-t_1) F(t_1-t_2) k(t_1-{\tau_1})  k(t_2-{\tau_2}) \hat K_1 \hat K_2.
\end{equation}
This on its own is not yet a feature stemming only from the imperfection of the clocks, as even for perfect clocks these terms appear if the support of $k(t-\tau_1)$ and $k(t-\tau_2)$ overlap.
However, if we choose $k(t-\tau_1)$ and $k(t-\tau_2)$ such that their supports do not overlap, then in the ideal case this term should not contribute. 
An illustration of this case is shown in \cref{fig:two_measurements} for perfect and imperfect clocks. 
As $F(t_1-t_2)$ approximates but is not equal, to Heaviside's theta function $\Theta(t_1-t_2)$, the term in \cref{acausal} is not zero even if $t_1-t_2 < 0$. 

We conclude that, if the clocks are no longer perfect, and events are defined according to the \GLMshort approach, then there is a limitation to the notion of time ordering of events, stemming from the time-non-locality of the evolution equation. 
This effect is another example of the inequivalence between the \HSLshort and \GLMshort approaches.

As discussed above, the \GLMshort definition of event corresponds to the description that an external observer would give of an internal experimenter looking at a clock and, conditioned on measuring a certain time, performing a measurement on a quantum system. 
From this perspective, these considerations might be relevant to the study of indefinite causal structure in the sense of process matrices and related operational frameworks \cite{Oreshkov_2012, Baumann_2022, Oreshkov_2019}. 
The causal inequalities introduced in the field of indefinite causal orders would provide an interesting test for how strong the limitations on time-ordering of events are that stem from a non-ideal clock. We currently do not know if this evolution equation would lead to a violation of such inequalities.

\subsubsection{Comparison with the non-ideal \HSL approach}

It is clear from the analysis above that, away from the limit of ideal clocks, the \HSLshort and \GLMshort approaches are physically different. 
We now propose an operational interpretation of each, accounting for this discrepancy.

In the \HSLshort approach, we interpret events in the following way: there is an external experimenter who can perform measurements relative to the clock on the system. 
The joint state of the clock and the system is ``frozen''.
Moreover, the order of the measurements of the experimenter is not bound by the clock, that is, they can apply the measurements in any order they like. 
This is not only a feature of \HSLshort but also in ordinary quantum mechanics where it is possible to measure an operator $\hat{A}_S(t)$ at any time $t'$ by evolving the corresponding operator to compensate for the time difference. The state of the system is given by \cref{hslhistorystate} and the operations are given by \cref{eq:HSLop} (suitably modifying each of them to the case of non-ideal clocks). Interestingly, for an imperfect yet sufficiently ``good'' clock (see \cite{Hoehn_2021} for the exact conditions), this description is mathematically equivalent, by means of conditioning on the clock to be on a state $\ket{\phi_t}_C$, to the standard quantum mechanical description of a system evolving in an external background time. 
Specifically,
\begin{align}\label{examplee}
\brat{t} &\hat F_{CS}(\tau_2) \hat G_{CS}(\tau_1) \ket{\Psi}_{CS} = e^{ -i \hat H_S (t-\tau_2)} \hat F_S e^{ -i \hat H_S (\tau_2-\tau_1)}\hat G_S e^{ -i \hat H_S \tau_1} \ket{\psi_0}_S.
\end{align}
However, the probe used to perform such measurements and the order of events in this approach are given \textit{externally} (for instance using an external clock) and are, operationally, unrelated to the time $\tau$ in \cref{eq:constraint} (or $\tau_1$ and $\tau_2$ in \cref{examplee}). 
Therefore, the internal spacetime interpretation of ``events at a time $\tau$'' in the \HSLshort approach is a \textit{reconstruction} based on observations ``from the outside''.

Importantly, the resources of the clock do not play any role in deciding the order in which the operations $\hat F_{CS}(\tau_2)$ and $\hat G_{CS}(\tau_1)$ (or equivalently $\hat F_S$ and $\hat G_S$) are applied. 
It is as if the external experimenter had access to a continuous set of operators labelled by a parameter $\tau$, and she had the power to choose unambiguously specific operators from the set and apply them in any order she wants. 
In this sense, she has infinite resources, even if these are not provided by a temporal reference frame. 

On the other hand, the \GLMshort approach can be interpreted as the external description of an internal observer who looks at a clock and triggers a measurement interaction on the system when they see the clock showing a certain time. 
This action constitutes the event. 
By contracting the solution to \cref{eq:const_imperfect_glm} with $\bra{\phi_t}$, we obtain \cref{eq:imperfect_evolution}, which we can interpret as the description of events ``as seen'' from the perspective of the quantum clock $C$ (although in a ``purified'' sense). 
In this case, the time order between events is determined \textit{dynamically}, by the interaction between the clock and the systems. This is clearly operationally different from the notion of event in the \HSLshort approach. 
Therefore, the inequivalence between the two frameworks is due to a different implicit notion of what an ``event at time $t$'' means operationally.

In particular, the resources of the clock become all important in the \GLMshort approach. 
For clocks with finite resources, the measurement interaction induces time-non-local effects, which imply that the evolution of the system ``as seen'' by the clock is in general not unitary and that there is a fundamental limitation to the operational time order of events, as seen in subsection \ref{indeftemp}. 
(Non-unitarity can also arise in the $\sigma = \mathbb{R}$ case, see \cite{Smith_Ahmadi_2019,Paiva_2022,Paiva2022,Rijavec2023}.) 
Moreover, the denominator of the probability rule, \cref{1prob} becomes non-trivial, which is a departure from the Born rule of ordinary quantum mechanics. 

To translate between one formalism and the other, in \cref{appendix:glm_to_hsl} we show how to compute \GLMshort probabilities in terms of \HSLshort conditional probabilities, where evolution is unitary. 
Not surprisingly, the resulting invariant observables in the \HSLshort approach have time-non-local support. 
Thus, the \GLMshort approach can be interpreted in terms of states which are solutions to the constraint given by the \HSLshort approach, but with a restriction on the available class of operators. 
This restriction, which also implies a form of time-non-locality, is a consequence of the limited resources of the clock. 
Finally, to illustrate the difference between the two approaches in an extreme case, imagine that the clock has no dynamics, that is, set $\hat{H}_C =0$. 
In the \HSLshort approach, it is still meaningful to talk about measurements. 
Even if the state of the system does not evolve relative to the clock, the operator $\hat{F}_{CS}(\tau)$ will modify the state in a non-trivial way. 
In contrast, setting $\hat{H}_C =0$ in the \GLMshort approach makes measurements meaningless, as a working clock is necessary to trigger the interaction by which the ancilla $A$ gathers information about the measurement outcome. 

\section{Non-unitarity and discreteness of time}\label{sec:discrete}
In the above discussion, we tacitly assumed that time, and consequently time evolution, is continuous. 
This, and the assumption that the clock has a bounded spectrum, led us to conclude that, when measurements are modelled dynamically, as in the \GLMshort approach, the evolution equation governing the dynamics is time-non-local, in general leading to a non-unitary evolution. 

In this Section, we drop the assumption that time evolution is continuous while keeping the resources of the clock bounded. 
We construct a model for the \GLMshort approach to measurements and show that it is consistent both with unitary evolution and bounded resources. 
A similar model was previously constructed in \cite{Mondragon_2007}. 

If we consider only a discrete set of times, we are only interested in the system being invariant under discrete time evolution, rather than an infinitesimal time evolution generated by a Hamiltonian. 
We define the discrete time evolution operator by
\begin{equation}
    \hat U = \sum_{k} \ketbra{t_{k+1}}{t_k}_C \otimes \hat U^{(k)}_{SA},
\end{equation}
with $\{\ket{t_k}| t_{k} = \frac{k \pi}{E}, \ k \in \mathbb{Z} \}$
being an orthonormal basis spanning the Hilbert space of a clock with spectrum $[-E,E]$, such that $\ket{t_{k+1}} = e^{- i \hat H_C \frac{\pi}{E}} \ket{t_{k}}$.
Note that making time discrete is crucial here, as then such an orthonormal basis exits for clocks with bounded energy spectrum. This is not possible if time is continuous, i.e., $k$ can take any value in $\mathbb{R}$.
The unitary operators $\hat U^{(k)}_{SA}$ act on the system $S$ and an ancilla $A$. 
They model a general measurement interaction in which the ancilla records the result of a measurement performed on the system at a specified time. 
In this way, the measurement process forms part of the dynamics, making this model a discrete version of the \GLMshort approach. Note, however, that this model is not obtained by directly discretising the continuous-time model of section \cref{nonidealGLM}. 

As noted above, the analogue of the Hamiltonian constraint in the discrete case is invariance under finite time evolution. 
Therefore, we demand that the global state $\ket{\Psi}_{CSA}$ satisfies
\begin{equation}\label{discreteconstraint}
\hat U \ket{\Psi}_{CSA} = \ket{\Psi}_{CSA}.
\end{equation}
Note that this constraint is weaker than the constraint of \cref{eq:const_imperfect_glm}, as we only require that the system is invariant under the discrete evolution.

As we show in \cref{appendix:discrete_time}, the solution to \cref{discreteconstraint} reads
\begin{equation}\label{eq:sol_discrete}
    \sum_{k} \ket{t_k}_C \otimes \ket{\psi (t_k)}_{SA},
\end{equation}
with
\begin{equation}  
\ket{\psi(t_{k+1})}_{SA} = \prod_{j = 0}^k \hat U^{(j)}_{SA} \ket{\psi(t_{0})}_{SA}.
\end{equation}
Note that, in \cref{eq:sol_discrete}, the evolution of the system with respect to the clock is completely unitary. 

Based on this observation, we arrive at the following statement: if, (\textit{i}) every physical clock has a bounded energy spectrum, (\textit{ii}) measurements of a system at a certain time couple the clock to the system, and (\textit{iii}) we demand that the system evolves locally and unitarily with respect to the clock, then our analysis provides evidence that the underlying time, understood operationally as what a clock can show, has to be discrete. Note, however, that this model does not exclude the possibility that discrete time evolution could be non-unitary.

\section{Discussion}
We have analysed two recent approaches to measurements in the Page-Wootters formalism -- the TO approach and the PM approach. 
Each approach solves \kuch criticisms differently: the TO approach defines measurements in terms of time-translation invariant operators, and the PM approach includes explicitly a dynamical description of the measurement process. 
We have shown that, whereas the two approaches produce the same predictions in the limit of ideal clocks, their predictions differ in the non-ideal case. 
We have compared the two solutions to reach a better operational understanding of measurement events in the ``timeless'' approach to quantum mechanics. 

The TO approach, closely related to Dirac quantization, is based on time translation invariant operators applied to a static state. 
The operators are applied in an order that is chosen externally, using resources other than the clock's to distinguish between them. 
The resulting picture can be ``internalized'' and reinterpreted as ordinary quantum mechanics with external background time. 
In this reconstruction of events and time evolution, we have infinite resources to distinguish between different operations and determine their order, even if the clock is not ideal.   

The PM approach describes measurements dynamically, and the resources of the clock are important to determine the order in which measurement interactions happen. 
It can be interpreted as the external description of an internal observer who measures a system when she ``sees'' her clock reading a certain time. 
Note that both the internal and external descriptions are relative to the same clock. 
However, while the measurement event consists of a detector ``click'' in the internal description, it takes the form of a correlation between the system, the clock and an ancilla in the external description. 
When the clocks are far from the ideal case, time evolution looks very different to the unitary evolution of ordinary quantum mechanics. 
Specifically, it leads to a time non-local evolution equation, which generically implies a non-unitary evolution. 
In this approach, the fact that the clock has finite resources becomes very important. 
Most strikingly, the lack of resources to distinguish perfectly between different clock states leads to fundamental limitations to the notion of time ordering.

Therefore, the reconstruction of measurement events in the TO approach coincides with the operational notion of ``looking'' at a clock and measuring the system only in the limit where clocks can be treated as ideal. 
Intriguingly, even if both approaches are operationally different, they predict the same results in this limit. 
A possible explanation for this is that a clock with infinite resources in the \GLMshort approach is not affected by its coupling to the system and the ancilla, becoming effectively the external resource deciding the measurements in the \HSLshort approach.
However, a full explanation of this fact lies beyond the scope of this work.

Our analysis opens the question of how to properly describe the purification of measurements in timeless approaches to quantum theory.
Measurement interactions carry energy, and this has non-negligible consequences at the level of the constraint. 
The question of purification is important given the role of purification as a foundational principle for quantum mechanics \cite{Chiribella_2011}. 
Furthermore, it would be interesting to see how the analysis presented here extends to the case where we have multiple clocks describing a single system, as done in recent works on quantum reference frames, e.g. \cite{H_hn_2020, Castro_Ruiz_2020}.

In the non-ideal case, what are the implications of the \GLMshort model for spacetime? 
If time is defined operationally, in terms of what a clock measures, then a fundamental limit on our clock's resources would also imply a fundamental limit on our notion of time as we understand it classically. 
Considering the gravitational back reaction of a quantum clock on spacetime strengthens the operational point of view. 
In such scenarios, there are physical situations that fundamentally prohibit the measurement of time with more than a certain accuracy, limiting the resources of our clocks. 

This operational stance resonates with the analysis of Peres in \cite{Peres_1980}, who noted that: 
\begin{quote}
    It thus seems that the Schrödinger wave function $\psi(t)$, with its continuous time evolution given by $i \hbar \dot \psi = H \psi$, is an idealization rooted in classical theory. It is operationally ill-defined (except in the limiting case of stationary states) and should probably give way to a more complicated dynamical formalism, perhaps one non-local in time. Thus, in retrospect, the Hamiltonian approach to quantum physics carries the seeds of its own demise.
\end{quote}
The time-non-local evolution and the indefinite time order of measurement events explored here are instances of such a description.

Time non-locality and indefinite temporal order are, however, not the only possibilities. 
Returning to Peres' remarks, if continuous time evolution is operationally ill-defined, perhaps a discrete treatment of time would be the right thing to do. 
This resonates with our findings: in a timeless setting where the Hamiltonian evolution is generated by the relative degrees of freedom between a clock and the system, unitary evolution, definite time order and a clock with a bounded spectrum can be recovered if the emergent time model is discrete.

\section*{Acknowledgements}

\noindent 
L.H. acknowledges support from the Swiss National Science Foundation (SNSF) via project grant No. 200021\_188541 and the Quantum Center of ETH Zürich. E.C.-R. is supported by an ETH Zurich Postdoctoral Fellowship and acknowledges financial support from the Swiss National Science Foundation (SNSF) via the National Centers of Competence in Research QSIT and SwissMAP, as well as the project No. 200021\_188541. This publication was made possible through the support of the ID\# 61466 grant and ID\# 62312 grant from the John Templeton Foundation, as part of the project \href{http://www.qiss.fr}{‘The Quantum Information Structure of Spacetime’ (QISS)}. The opinions expressed in this publication are those of the authors and do not necessarily reflect the views of the John Templeton Foundation. It was also supported by the PEPR integrated project EPiQ ANR-22-PETQ-0007 part of Plan France 2030.

\bibliographystyle{quantum}

\begin{thebibliography}{10}

\bibitem{Wheeler_1978}
John~A. Wheeler.
\newblock ``The {\textquotedblleft}past{\textquotedblright} and the {\textquotedblleft}delayed-choice{\textquotedblright} double-slit experiment''.
\newblock In Mathematical Foundations of Quantum Theory.
\newblock \href{https://dx.doi.org/10.1016/b978-0-12-473250-6.50006-6}{Pages 9--48}.
\newblock Elsevier~(1978).

\bibitem{Page_1983}
Don~N. Page and William~K. Wootters.
\newblock ``Evolution without evolution: Dynamics described by stationary observables''.
\newblock \href{https://dx.doi.org/10.1103/physrevd.27.2885}{Physical Review D {\bf 27}, 2885--2892}~(1983).

\bibitem{Wootters_1984}
William~K. Wootters.
\newblock ``{``Time''} replaced by quantum correlations''.
\newblock \href{https://dx.doi.org/10.1007/bf02214098}{International Journal of Theoretical Physics {\bf 23}, 701--711}~(1984).

\bibitem{Kuchar_2011}
Karel~V. Kucha{\v{r}}.
\newblock ``Time and interpretations of quantum gravity''.
\newblock \href{https://dx.doi.org/10.1142/s0218271811019347}{International Journal of Modern Physics D {\bf 20}, 3--86}~(2011).

\bibitem{Unruh_1989}
William~G. Unruh and Robert~M. Wald.
\newblock ``Time and the interpretation of canonical quantum gravity''.
\newblock \href{https://dx.doi.org/https://doi.org/10.1103/PhysRevD.40.2598}{Physical Review D {\bf 40}, 2598--2614}~(1989).

\bibitem{Hoehn_2021}
Philipp~A. Höhn, Alexander R.~H. Smith, and Maximilian P.~E. Lock.
\newblock ``Trinity of relational quantum dynamics''.
\newblock \href{https://dx.doi.org/10.1103/physrevd.104.066001}{Physical Review D {\bf 104}, 066001}~(2021).

\bibitem{Dolby_2004}
Carl~E. Dolby.
\newblock ``The conditional probability interpretation of the {Hamiltonian} constraint''~(2004).
\newblock  \href{http://arxiv.org/abs/gr-qc/0406034}{arXiv:gr-qc/0406034}.

\bibitem{Dirac_2001}
P.~A.~M. Dirac.
\newblock ``Lectures on quantum mechanics''.
\newblock Dover Publications, Mineola, NY. ~(2001).

\bibitem{Henneaux_1992}
Marc Henneaux and Claudio Teitelboim.
\newblock ``Quantization of gauge systems''.
\newblock Princeton University, Princeton, N.J. ~(1992).

\bibitem{Bojowald_2011}
Martin Bojowald, Philipp~A. Höhn, and Artur Tsobanjan.
\newblock ``Effective approach to the problem of time: General features and examples''.
\newblock \href{https://dx.doi.org/https://doi.org/10.1103/PhysRevD.83.125023}{Physical Review D {\bf 83}, 125023}~(2011).

\bibitem{Bojowald_2011a}
Martin Bojowald, Philipp~A Höhn, and Artur Tsobanjan.
\newblock ``An effective approach to the problem of time''.
\newblock \href{https://dx.doi.org/10.1088/0264-9381/28/3/035006}{Classical and Quantum Gravity {\bf 28}, 035006}~(2011).

\bibitem{Gambini_2004}
Rodolfo Gambini, Rafael~A. Porto, and Jorge Pullin.
\newblock ``A relational solution to the problem of time in quantum mechanics and quantum gravity: a fundamental mechanism for quantum decoherence''.
\newblock \href{https://dx.doi.org/10.1088/1367-2630/6/1/045}{New Journal of Physics {\bf 6}, 45--45}~(2004).

\bibitem{Gambini_2009}
Rodolfo Gambini, Rafael~A. Porto, Jorge Pullin, and Sebasti{\'{a}}n Torterolo.
\newblock ``Conditional probabilities with {Dirac} observables and the problem of time in quantum gravity''.
\newblock \href{https://dx.doi.org/https://doi.org/10.1103/PhysRevD.79.041501}{Physical Review D {\bf 79}, 041501}~(2009).

\bibitem{Rovelli_1990}
Carlo Rovelli.
\newblock ``Quantum mechanics without time: A model''.
\newblock \href{https://dx.doi.org/https://doi.org/10.1103/PhysRevD.42.2638}{Physical Review D {\bf 42}, 2638--2646}~(1990).

\bibitem{Rovelli_2014}
Carlo Rovelli and Francesca Vidotto.
\newblock ``Covariant loop quantum gravity''.
\newblock \href{https://dx.doi.org/https://doi.org/10.1017/CBO9781107706910}{Cambridge University Press}. ~(2014).

\bibitem{RovelliC_2009}
Carlo Rovelli.
\newblock ``{Forget time}''~(2009).
\newblock  \href{http://arxiv.org/abs/0903.3832}{arXiv:0903.3832}.

\bibitem{Giovannetti_2015}
Vittorio Giovannetti, Seth Lloyd, and Lorenzo Maccone.
\newblock ``Quantum time''.
\newblock \href{https://dx.doi.org/10.1103/physrevd.92.045033}{Physical Review D {\bf 92}, 045033}~(2015).

\bibitem{Hellmann_2007}
Frank Hellmann, Mauricio Mondragon, Alejandro Perez, and Carlo Rovelli.
\newblock ``Multiple-event probability in general-relativistic quantum mechanics''.
\newblock \href{https://dx.doi.org/https://doi.org/10.1103/PhysRevD.75.084033}{Physical Review D {\bf 75}, 084033}~(2007).

\bibitem{Mondragon_2007}
Mauricio Mondragon, Alejandro Perez, and Carlo Rovelli.
\newblock ``Multiple-event probability in general-relativistic quantum mechanics. {II}. a discrete model''.
\newblock \href{https://dx.doi.org/https://doi.org/10.1103/PhysRevD.76.064005}{Physical Review D {\bf 76}, 064005}~(2007).

\bibitem{Aharanov1961}
Y.~Aharonov and D.~Bohm.
\newblock ``Time in the quantum theory and the uncertainty relation for time and energy''.
\newblock \href{https://dx.doi.org/10.1103/PhysRev.122.1649}{Phys. Rev. {\bf 122}, 1649--1658}~(1961).

\bibitem{Castro_Ruiz_2020}
Esteban Castro-Ruiz, Flaminia Giacomini, Alessio Belenchia, and {\v{C}}aslav Brukner.
\newblock ``Quantum clocks and the temporal localisability of events in the presence of gravitating quantum systems''.
\newblock \href{https://dx.doi.org/https://doi.org/10.1038/s41467-020-16013-1}{Nature Communications {\bf 11}, 2672}~(2020).

\bibitem{Giacomini_2019}
Flaminia Giacomini, Esteban Castro-Ruiz, and {\v{C}}aslav Brukner.
\newblock ``Quantum mechanics and the covariance of physical laws in quantum reference frames''.
\newblock \href{https://dx.doi.org/https://doi.org/10.1038/s41467-018-08155-0}{Nature Communications {\bf 10}, 494}~(2019).

\bibitem{H_hn_2012}
Philipp~A. Höhn, Em{\'{\i}}lia Kubalov{\'{a}}, and Artur Tsobanjan.
\newblock ``Effective relational dynamics of a nonintegrable cosmological model''.
\newblock \href{https://dx.doi.org/https://doi.org/10.1103/PhysRevD.86.065014}{Physical Review D {\bf 86}, 065014}~(2012).

\bibitem{H_hn_2020}
Philipp~A. Höhn and Augustin Vanrietvelde.
\newblock ``How to switch between relational quantum clocks''.
\newblock \href{https://dx.doi.org/10.1088/1367-2630/abd1ac}{New Journal of Physics {\bf 22}, 123048}~(2020).

\bibitem{H_hn_2021}
Philipp~A. Höhn, Alexander R.~H. Smith, and Maximilian P.~E. Lock.
\newblock ``Equivalence of approaches to relational quantum dynamics in relativistic settings''.
\newblock \href{https://dx.doi.org/10.3389/fphy.2021.587083}{Frontiers in Physics {\bf 9}, 587083}~(2021).

\bibitem{Giacomini_2021}
Flaminia Giacomini.
\newblock ``Spacetime quantum reference frames and superpositions of proper times''.
\newblock \href{https://dx.doi.org/https://doi.org/10.22331/q-2021-07-22-508}{Quantum {\bf 5}, 508}~(2021).

\bibitem{Loveridge_2019}
Leon Loveridge and Takayuki Miyadera.
\newblock ``Relative quantum time''.
\newblock \href{https://dx.doi.org/https://doi.org/10.1007/s10701-019-00268-w}{Foundations of Physics {\bf 49}, 549--560}~(2019).

\bibitem{Hoehn_2019}
Philipp Höhn.
\newblock ``Switching internal times and a new perspective on the `wave function of the universe'''.
\newblock \href{https://dx.doi.org/10.3390/universe5050116}{Universe {\bf 5}, 116}~(2019).

\bibitem{Bronstein_2011}
Matvei Bronstein.
\newblock ``Republication of: Quantum theory of weak gravitational fields''.
\newblock \href{https://dx.doi.org/https://doi.org/10.1007/s10714-011-1285-4}{General Relativity and Gravitation {\bf 44}, 267--283}~(2011).

\bibitem{Castro_Ruiz_2017}
Esteban~Castro Ruiz, Flaminia Giacomini, and {\v{C}}aslav Brukner.
\newblock ``Entanglement of quantum clocks through gravity''.
\newblock \href{https://dx.doi.org/https://doi.org/10.1073/pnas.1616427114}{Proceedings of the National Academy of Sciences {\bf 114}, E2303--E2309}~(2017).

\bibitem{Smith_Ahmadi_2019}
Alexander R.~H. Smith and Mehdi Ahmadi.
\newblock ``Quantizing time: Interacting clocks and systems''.
\newblock \href{https://dx.doi.org/https://doi.org/10.22331/q-2019-07-08-160}{Quantum {\bf 3}, 160}~(2019).

\bibitem{de_la_Hamette_2021}
Anne-Catherine de~la Hamette, Thomas~D. Galley, Philipp~A. Hoehn, Leon Loveridge, and Markus~P. Mueller.
\newblock ``Perspective-neutral approach to quantum frame covariance for general symmetry groups''~(2021).
\newblock  \href{http://arxiv.org/abs/2110.13824}{arXiv:2110.13824}.

\bibitem{Oreshkov_2012}
Ognyan Oreshkov, Fabio Costa, and {\v{C}}aslav Brukner.
\newblock ``Quantum correlations with no causal order''.
\newblock \href{https://dx.doi.org/https://doi.org/10.1038/ncomms2076}{Nature Communications {\bf 3}, 1092}~(2012).

\bibitem{Baumann_2022}
Veronika Baumann, Marius Krumm, Philippe~Allard Gu{\'{e}}rin, and {\v{C}}aslav Brukner.
\newblock ``Noncausal page-wootters circuits''.
\newblock \href{https://dx.doi.org/https://doi.org/10.1103/PhysRevResearch.4.013180}{Physical Review Research {\bf 4}, 013180}~(2022).

\bibitem{Oreshkov_2019}
Ognyan Oreshkov.
\newblock ``Time-delocalized quantum subsystems and operations: on the existence of processes with indefinite causal structure in quantum mechanics''.
\newblock \href{https://dx.doi.org/https://doi.org/10.22331/q-2019-12-02-206}{Quantum {\bf 3}, 206}~(2019).

\bibitem{Ahmadi_2013}
Mehdi Ahmadi, David Jennings, and Terry Rudolph.
\newblock ``The wigner{\textendash}araki{\textendash}yanase theorem and the quantum resource theory of asymmetry''.
\newblock \href{https://dx.doi.org/10.1088/1367-2630/15/1/013057}{New Journal of Physics {\bf 15}, 013057}~(2013).

\bibitem{Bartlett_2007}
Stephen~D. Bartlett, Terry Rudolph, and Robert~W. Spekkens.
\newblock ``Reference frames, superselection rules, and quantum information''.
\newblock \href{https://dx.doi.org/https://doi.org/10.1103/RevModPhys.79.555}{Reviews of Modern Physics {\bf 79}, 555--609}~(2007).

\bibitem{baumann2021generalized}
Veronika Baumann, Flavio Del~Santo, Alexander R.~H. Smith, Flaminia Giacomini, Esteban Castro-Ruiz, and Caslav Brukner.
\newblock ``Generalized probability rules from a timeless formulation of wigner's friend scenarios''.
\newblock \href{https://dx.doi.org/https://doi.org/10.22331/q-2021-08-16-524}{Quantum {\bf 5}, 524}~(2021).

\bibitem{Trassinelli2022}
M.~Trassinelli.
\newblock ``Conditional probabilities of measurements, quantum time, and the wigner's-friend case''.
\newblock \href{https://dx.doi.org/10.1103/PhysRevA.105.032213}{Phys. Rev. A {\bf 105}, 032213}~(2022).

\bibitem{Paiva_2022}
Ismael~L. Paiva, Amit Te'eni, Bar~Y. Peled, Eliahu Cohen, and Yakir Aharonov.
\newblock ``Non-inertial quantum clock frames lead to non-hermitian dynamics''.
\newblock \href{https://dx.doi.org/https://doi.org/10.1038/s42005-022-01081-0}{Communications Physics {\bf 5}, 298}~(2022).

\bibitem{Paiva2022}
Ismael~L. Paiva, Augusto~C. Lobo, and Eliahu Cohen.
\newblock ``Flow of time during energy measurements and the resulting time-energy uncertainty relations''.
\newblock \href{https://dx.doi.org/10.22331/q-2022-04-07-683}{Quantum {\bf 6}, 683}~(2022).

\bibitem{Rijavec2023}
Simone Rijavec.
\newblock ``Robustness of the page-wootters construction across different pictures, states of the universe, and system-clock interactions''.
\newblock \href{https://dx.doi.org/10.1103/PhysRevD.108.063507}{Phys. Rev. D {\bf 108}, 063507}~(2023).

\bibitem{Chiribella_2011}
Giulio Chiribella, Giacomo~Mauro D'Ariano, and Paolo Perinotti.
\newblock ``Informational derivation of quantum theory''.
\newblock \href{https://dx.doi.org/10.1103/physreva.84.012311}{Physical Review A {\bf 84}, 012311}~(2011).

\bibitem{Peres_1980}
Asher Peres.
\newblock ``Measurement of time by quantum clocks''.
\newblock \href{https://dx.doi.org/10.1119/1.12061}{American Journal of Physics {\bf 48}, 552--557}~(1980).

\bibitem{Marolf_2000}
Donald Marolf.
\newblock ``Group averaging and refined algebraic quantization: Where are we now?''~(2000).
\newblock  \href{http://arxiv.org/abs/gr-qc/0011112}{arXiv:gr-qc/0011112}.

\end{thebibliography}

\appendix

\section{Review of the Page-Wootters Formalism}
\noindent Here we give a pedagogical introduction to the Page-Wootters formalism \cite{Page_1983,Wootters_1984}.
The relational nature and reparametrization-invariance of general relativity suggest that the physical states $\ket{\Psi}$ describing a theory of quantum gravity should satisfy a Wheeler-DeWitt equation 
\begin{equation}
	\hat H \ket{\Psi} = 0,
	\label{cons1}
\end{equation} 
where $\hat H$ is the total \emph{Hamiltonian constraint}. 
An immediate consequence is that states evolving according to the Schrödinger equation
\begin{equation}
	i \frac{\partial}{\partial t} \hs = \hat H \hs = 0,
\end{equation} 
appear frozen in time. For this reason, theories satisfying \cref{cons1} are often called \emph{timeless}. 
Don Page and William Wootters showed that one can nevertheless recover the correct dynamics using a mechanism that is now known as Page-Wootters formalism. 
In the following, we give a concise summary of this approach.

In the Page-Wootters formalism, the evolution of a system $S$ is described by considering states $\hs$ not on the Hilbert space $\hal_S$ of the system, but on an extended Hilbert space $\hs \in \hal_C \tens \hal_S$ \footnote{Alternatively, one can introduce the Page-Wootters mechanism by considering a global Hilbert space $\mathcal{H}$, choosing one of the subsystems as a clock, and looking at the evolution of the system relative to this choice of clock.}. 
Here, $C$ is an ancilla system referred to as the clock. The clock is equipped with a canonical pair of time and energy coordinates $\hat T_C, \hat H_C$, satisfying 
\begin{equation}
	\label{comm1}
	[\hat T_C, \hat H_C] = i.
\end{equation}
The Hamiltonian constraint \cref{cons1} translates to 
\begin{equation}
	 \hat H\hs = 0, \quad \hat H = \hat H_C + \hat H_S \equiv \hat H_C\tens \mathbb{1} + \mathbb{1} \tens \hat H_S.
\end{equation}

Page and Wootters consider the case of an \emph{ideal clock} $C$, meaning $\hat T_C$ is a self-adjoint operator with generalized orthogonal eigenvectors $\ket{t}_C$, such that $\hat T_C \ket{t}_C = t \ket{t}_C$ and $\braket{t'}{t}_C = \delta(t-t')$. In that case, we can interpret $\ket{\psi(t)}_S = \bra{t}_C\ket{\Psi}_{CS}$ as the state of the system $S$ when clock $C$ shows time $t$. Indeed, the Hamiltonian constraint in \cref{cons1} recovers the correct dynamics for $\ket{\psi(t)}_S$, that is the Schrödinger equation,
\begin{equation}\label{eq:SG_emergent}
	\bra{t}_C \tot\ket{\Psi}_{CS} = 0 \quad \leftrightarrow\quad 
	i \dif{}{t} \ket{\psi(t)}_S = \hat H_S\ket{\psi(t)}_S.
\end{equation}
This follows from the commutation relation given in \cref{comm1} and the Stone-von Neumann theorem, which together imply that $\hat H_C$ is represented as a differential operator $\hat H_C \ket{\psi(t)}_S = -i \del_t \ket{\psi(t)}_S$. In the $\ket{t}_C$ basis, the state $\ket\Psi$ then takes the form 
\begin{equation}
	\ket\Psi = \int dt \ket{t}_C \tens \ket{\psi(t)}_S,
	\label{hstate}
\end{equation} 
which is why it is often referred to as the history state. 

One-time measurements of an observable $\hat O_S = \sum_{a} O_a \ketbra{a}{a}$ at time $\tau$ can be defined in the Page-Wootters mechanism by first conditioning on finding the clock in state $\ket\tau_C$: 
The probability to find the system $S$ in state $\ket{a}_S$ at time $\tau$ is thus given by 
\begin{equation}	
	P(a|\tau) = \frac{\bra{\Psi}\ \ketbra{\tau}{\tau}_C\tens \ketbra{a}{a}_S \ \ket{\Psi}_{CS}}{\bra\Psi \ (\ketbra{\tau}{\tau}_C) \ \ket \Psi_{CS}} = \frac{\big \vert {\bra{\psi(\tau)}\ket{a}_S}\big \vert^2}{\big \| \ket{\psi(\tau)}\big\|^2}.
 \label{eq:onetimemes}
\end{equation} 
This probability is in agreement with Born's rule and thus recovers the correct dynamics of the system $S$ for one-time measurements; however, for consecutive measurements, one should be more careful, as we will see below. 
The issues with consecutive measurements give rise to the different approaches to measurements in the Page-Wootters mechanism discussed in this paper. 

\label{appendix:page_wooters}

\section{Review of \kuch
 criticisms}\label{appendix:kuchar}
While \cref{eq:onetimemes} is in agreement with Born's rule for one-point measurements, it immediately leads to issues when multiple subsequent measurements are performed. These issues were first raised by \kuchar \cite{Kuchar_2011} as follows:

\noindent \textbf{1. Violation of constraints}: In the Page-Wootters mechanism, the measurement of an observable $\hat{O}_S$ at time $\tau$ is described by a projective operator
\begin{equation}
	\hat \Pi_{CS}^{a,\tau} = \ketbra{\tau}{\tau}_C \otimes  \ketbra{a}{a}_S 
\end{equation} 
acting on $\hs_{CS}$, where $\ket{a}_{S}$ is an eigenstate of $\hat O_S$. 
However, this operator, in general, does not commute with the constraint $\tot$.
Indeed, as the constraint is the total Hamiltonian, any operator commuting with it must be stationary, while there is obvious experimental evidence that the above projector is time-dependent. 
In that case, applying the measurement operator throws $\ket\Psi$ out of the constrained Hilbert space. \\
\textbf{2. Wrong propagators}: As a consequence, the Page-Wootters mechanism does not give the correct propagators for multiple subsequent measurements. 
For example, the dynamical question "If the system is in state $\ket{a}_S$ at time $\tau$, what is the probability to find it in state $\ket{b}_S$ at time $\tau'>\tau$?" cannot be answered by applying the conditional probability in equation \eqref{eq:onetimemes} twice. 
This would yield 
\begin{equation}
	P((b|\tau')|(a|\tau)) = \frac{\bra\Psi \hat \Pi_{CS}^{a,\tau} \hat \Pi_{CS}^{b,\tau'} \hat \Pi_{CS}^{a,\tau} \ket \Psi}{\bra\Psi \hat \Pi_{CS}^{a,\tau} \ket\Psi} = \abs{\delta(\tau'-\tau)}^2\abs{\bra{b}\ket{a}},
\end{equation}
which is non-zero only for $\tau' = \tau$ and not in accordance with Born's rule. 

\kuch first criticism concerns the fact that projective operators and the constraint $H$ do not commute in general. 
To resolve this issue, it seems natural to either modify the constraint (and therefore the history state $\hs$) or change the way operators are represented. 
These two approaches exactly correspond to the formalisms studied in the main body of this paper, the \GLM \cite{Giovannetti_2015} approach and \HSL \cite{Hoehn_2021} approach respectively. 

\section{Further details of the \GLM and \HSL approaches}\label{appendix:perfect_clocks}
\subsection{Review of ideal \GLM approach}\label{appendix:perfect_clocks:glm}
Giovannetti \textit{et al.}~\cite{Giovannetti_2015} proposed an extension to the Page-Wootters mechanism which overcomes the deficiencies raised by \kuchar. 
They do so by associating a Hamiltonian to each measurement, and including it in the constraint as follows: Let us consider an additional purifying ancilla $A$, which allows the description of a projective measurement at time $\tau$ through its Kraus representation as a unitary mapping 
\begin{equation}
	\hat V: \ket{\psi(\tau)}_S \tens \ket{r}_A \mapsto \sum_{a}\hat {K}_S^a\ket{\psi(\tau)}_S \tens\ket{a}_A
	\label{mes2}
\end{equation} 
acting on $\hal_S\tens\hal_A$, where the ancilla $A$ is initialized in a ready state $\ket r_A$ and stores the measurement outcome.
Here, $\hat {K}_S^a$ are Kraus operators satisfying $\sum_a  (\hat{K}_S^a)^\dagger \hat {K}_S^a = \mathbb{1}_S.$ 
Let $\hat H_{SA}$ be the Hamiltonian responsible for the measurement unitary $\hat V$, i.e. $\hat V = e^{-i \hat H_{SA}}$. Giovannetti \textit{et al.}~\cite{Giovannetti_2015} consider the case where $\hat H_A$ contributes to the total Hamiltonian constraint in the form 
\begin{equation}
	\label{eq:ideal_tot_H}
	\hat H = \hat H_C + \hat H_S + \delta(\hat T_C-\tau)\tens \hat H_{SA},
\end{equation} 
where $\delta(\hat T_C-\tau)\tens \hat H_{SA}$ acts on $\hal_C \tens \hal_S\tens\hal_A$. 
The solution to the constraint $\hat{H}\ket{\Psi}=0$, where $\hat{H}$ is given by \cref{eq:ideal_tot_H}, can be found by coherent group averaging \cite{Marolf_2000}. 
This yields the history state
\begin{align}
	\ket\Psi &= \int_{-\infty}^{\tau} dt \ket{t}_C \tens \ket{\psi(t)}_S \tens \ket{r}_A \\&+ \int_{\tau}^{\infty} dt  \ket{t}_C \tens \sum_{a} \hat U_S(t-\tau) \hat K^{a}_S\ket{\psi(\tau)}_S \tens \ket{a}_A.
	\label{app:1meshs}
\end{align}
The probability that outcome $a$ will be registered by the memory $A$ at a given time $t$ is given by projecting onto the measurement ancilla
\begin{equation}
	\label{app:1prob}
	P(a|t) = \frac{\bra\Psi \ketbra{a}{a}_A \otimes \ketbra{t}{t}_C  \hs}{\bra\Psi \ketbra{t}{t}_C \hs} .
\end{equation} 
By construction, \kuch first criticism is resolved because the measurement projector now acts on the measurement ancilla and not on the system $S$. 
In the special case of projective, non-degenerate von Neumann measurements, the Kraus operators are $\hat K^{a}_S = \ketbra{a}{a}_S$ and the probability reduces to
\begin{equation}
	P(a|\tau)_t = \big \|{\bra{a}_S \ket{\psi(t)}_S}\big \|^2,
\end{equation}
in accordance with the Born rule if $t \geq \tau$. 

In contrast to the original Page-Wootters proposal, the \GLMshort approach allows the discussion of multiple subsequent measurements. Indeed, \cref{eq:ideal_tot_H} can be readily extended by a second measurement ancilla $A'$ 
\begin{equation}
		\tot = \hat H_C + \hat H_S + \delta(\hat T_C-\tau)\tens \hat H_{SA} +\delta(\hat T_C-\tau')\tens \hat H_{SA'}.
		\label{totOper}
\end{equation} 
Group averaging results in the history state 
\begin{align}
	\label{GLMstate}
	\ket\Psi &= \int_{-\infty}^{\tau} dt \ket{t}_C \tens \ket{\psi(t)}_S \tens \ket{r}_A \tens \ket{r}_{A'}\\
    &+\int_{\tau}^{\tau'} dt  \ket{t}_C \tens \sum_{a} \hat U_S(t-\tau)\hat {K}_S^a\ket{\psi(\tau)}_S \tens \ket{a}_A \tens \ket{r}_{A'} \\
    &+ \int_{\tau'}^{\infty}
	dt \ket{t}_C \tens \sum_{a,b} \hat U_S(t-\tau')\hat K^{b}_S \hat U_S(\tau'-\tau)\hat {K}_S^a\ket{\psi(\tau)}_S \tens \ket{a}_A \tens \ket{b}_{A'}.
	\label{2meshs}
\end{align}
It is important to note that the probability $P(a|t)$ in \cref{app:1prob} is not affected by the presence of the second ancilla. 
However, the probability of observing the measurement result $\ket{b}_{A'}$ at time $\tau'$ conditioned on outcome $\ket{a}_{A}$ at time $\tau$ can now be expressed as \begin{equation}
\label{condGLM}
P\left((b|\tau')|(a|\tau)\right)_t = \frac{\big \|{\bra{t}_C\tens \bra{b}_{A'}\tens \bra{a}_A\hs \big \|}^2}{\big \|{(\bra{t}_C\tens\bra{a}_A\hs\big \|}^2},
\end{equation} 
which results in 
\begin{equation}
	P\left((b|\tau')|(a|\tau)\right)_t = \frac{\norm{ \hat \Pi^{b}_{S}  \hat U(\tau'-\tau) \hat \Pi^{a}_{S} \ket{\psi(\tau)}}^2}{\norm{\hat \Pi^{a}_{S}\ket{\psi(\tau)}}^2}.
\end{equation}
if $t \geq \tau' > \tau$.
This is the correct expression for two-point measurements, in agreement with ordinary quantum mechanics. 
\kuch second criticism is thus resolved as well.

In \cref{app:1meshs,2meshs}, the measurement outcome is entirely encoded in the record kept by the measurement ancillas. 
From a philosophical viewpoint, Giovannetti \textit{et al.}~\cite{Giovannetti_2015} therefore adopt Wheeler’s operationalist stance \cite{Wheeler_1978} that "the past has no existence except as it is recorded in the present". 

\subsection{Review of the \HSL approach}

Höhn \textit{et al.}~\cite{Hoehn_2021} propose a different extension of the Page-Wootters mechanism, which we call the \HSL approach or \HSLshort approach. 
As in the Page-Wootters formalism evolution of $S$ is described by a time-independent state $\ket\Psi$ solving the constraint 
\begin{equation}\label{eq:constraint}
\tot \ket{\Psi} = 0, \quad \tot = \hat H_C + \hat H_S.
\end{equation} 
The constraint once again recovers the Schrödinger equation for $\ket{\psi(t)}_S = \bra{t}_C\ket{\Psi}_S$. 
However, the requirement \cref{eq:constraint} results in a different representation of operators acting on $\hs$: An operator $f_S$ acting on $\ket{\psi(\tau)}_S$ is now represented by the incoherent group twirl
\begin{equation}
\label{app:HSLop}
\hat F_{SC}(\tau) =  \int_{\mathbb R} dt \, \ e^{- i t\hat H_C} \, \ketbra{\tau}{\tau}_C \, e^{ i t\hat H_C}\tens e^{- i t\hat H_S}\, \hat F_S \, e^{ i t\hat H_S}
\end{equation} 
acting on $\hs$, where again $\hat U(t) = e^{-i\tot t} $ is the free time evolution. By construction, $\hat F_{SC}(\tau)$ is invariant under $\hat U(t)$ and as a result, $[\hat F_{SC}(\tau),\tot] = 0$, resolving \kuch first criticism. 

The states $\ket{\Psi}$, solutions to the constraint \cref{eq:constraint} are normalized with respect to the so-called \emph{physical} scalar product $\bra{\cdot}\ket{\cdot}_{\text{phys}}$ \cite{Hoehn_2021,Smith_Ahmadi_2019}, which is defined as 
\begin{equation}
\bra{\Psi}\ket{\Phi}_{\text{phys}} = \bra{\psi(t)}_S \otimes \bra{t}_C \left(\int ds \ \hat U(s) \right) \ket{t }_C \otimes \ket{\phi(t)}_S,
\end{equation} 
for any time $t$. It is straightforward to verify that this scalar product is well-defined and in particular independent of the choice of $t$. 

The probability to find the system in state $\ket{a}_S$ at time $\tau$ is defined by 
\begin{equation}
	\label{app:mesHSL}
	P(a|\tau) = \bra{\Psi} \hat \Pi^{a}_{CS}(\tau) \ket{\Psi}_{\text{phys}},
\end{equation} 
where $\hat \Pi^{a}_{CS}(\tau) =  \int_{\mathbb R} dt \, \ e^{- i t\hat H_C} \, \ketbra{\tau}{\tau}_C \, e^{ i t\hat H_C}\tens e^{- i t\hat H_S}  \ketbra{a}{a}_S e^{i t\hat H_S}$. This coincides with the TO probability introduced in \cref{sec:HSL}.
This indeed gives the correct measurement statistics, as readily verified (setting $t=0$):
\begin{align}
	P(a|\tau) &= \bra{\Psi} \left[\int dt \ e^{- i t(\hat H_C + \hat H_S)}(\ketbra{\tau}{\tau}_C \tens \hat \Pi^{a}_S) e^{i t(\hat H_C + \hat H_S)} \right]\ket\Psi_{\text{phys}} \\ 
	&= \bra{\psi(0)}_S \otimes \bra{0}_C \left[\int ds\ e^{- i s(\hat H_C + \hat H_S)} \int dt \  e^{- i t(\hat H_C + \hat H_S)} (\ket{\tau}\bra{\tau}_C \tens \hat \Pi_S^{a}) \hat e^{i t(\hat H_C + \hat H_S)} \right] \ket{0}_C \otimes \ket{\psi(0)}_S \\
	&= \ \bra{\psi(0)}_S \otimes \bra{0}_C \left[\int ds\  e^{- i (s-\tau)(\hat H_C + \hat H_S)} \tens  \Pi_S^{a} \right]\ket{\tau}_C \otimes \ket{\psi(\tau)}_S \\
	&= \bra{\psi(\tau)} \hat \Pi_S^{a} \ket{\psi(\tau)}_S,
\end{align}
in agreement with Born's rule. 

Because measurements in the \HSLshort picture commute with $\tot$, they do not throw $\hs$ out of the constrained Hilbert space. Consequently, \cref{app:mesHSL} can be readily extended to more than one measurement. Furthermore, one can define conditional probabilities of the form 
\begin{equation}
	\label{app:condHSL}
	P((b|\tau')|(a|\tau)) = \frac{\bra{\Psi} \hat \Pi^{a}_{SC}(\tau) \hat \Pi^{b}_{SC}(\tau') \hat \Pi^{a}_{SC}(\tau) \ket\Psi_{\text{phys}}}{\bra\Psi \hat \Pi^{a}_{SC}(\tau) \ket \Psi_{\text{phys}}}.
\end{equation}
We will show in the next section that this expression indeed recovers the correct measurement statistics for two-point measurements by relating it to the \GLMshort approach. Hence the \HSLshort approach also resolves \kuch second criticism.

\section{Equivalence of predictions for perfect \GLMshort vs \HSLshort}\label{appendix:glm=hls}
\noindent The \GLMshort  and \HSLshort approach presented above are seemingly different from both a mathematical and philosophical perspective.
Giovanetti \textit{et al.} \cite{Giovannetti_2015} overcome \kuch criticisms by considering ancillas storing the measurement outcomes and by modifying the constraint operator $\tot$. Their procedure corresponds to an operationalist viewpoint. 
Höhn \textit{et al.} \cite{Hoehn_2021} overcome them by representing observables in a way that is invariant under the flow generated by the constraint, without modifying the constraint itself. 
We now show that these two \emph{a priori} different approaches are equivalent for multiple-time measurements in the case of perfect clocks, meaning they result in the same measurement outcome probabilities. 
Note that both approaches agree for single-time measurements already by construction.

To prove this, we want to show that the conditional probabilities \cref{condGLM,app:condHSL} of measuring System $S$ in state $\ket{b}_S$ at time $\tau'$, given that the system is measured to be in state $\ket{a}_S$ at time $\tau < \tau'$, are equal for both the \GLMshort and \HSLshort approach. 
To simplify the calculation, let us define the reduction map $R(\tau)$, which projects $\hs$ onto $\ket{\psi(\tau)}$, and its inverse on the constrained Hilbert space:
\begin{equation}
	\hat R(\tau) = \bra{\tau}_C, \quad \hat R^{-1}(\tau) = \int dt \ket{t}_C \hat U_S(t-\tau)
\end{equation}
Indeed, a change of variable yields 
\begin{equation}
	\hat R^{-1}(\tau) \hat R(\tau) \hs = \int dt \ \ketbra{t}{t}_C \hs = \hs.
\end{equation}
With this notation at hand, and using $\hat U^\dagger(t)\hs = \hs $ the operator acting on the constrained Hilbert space can be written as \begin{align}
	\hat F_{CS}(\tau) \hs &= 
	 \int dt \ \hat U(t) \ (\ket{\tau}\bra{\tau}\tens \hat f_S)\hs
	\\&= \int dt \ \ketbra{t}{\tau} \tens \hat U_S(t-\tau) \hat f_S \hs\\ &=( \hat R^{-1}(\tau) \hat f_S \hat R(\tau)) \hs.
	\label{reduction}
\end{align}
This allows us to expand the conditional probability \cref{app:condHSL} in terms of ordinary projection operators $\Pi^{a}_{S}, \Pi^{b}_{S}:$ 
\begin{align}
	P((b|\tau')|(a|\tau)) &= 
	\frac{\bra{\Psi} \hat \Pi^{a}_{SC}(\tau) \hat \Pi^{b}_{SC}(\tau') \hat \Pi^{a}_{SC}(\tau) \ket\Psi_{\text{phys}}}{\bra\Psi \hat \Pi^{b}_{SC}(\tau) \ket \Psi_{\text{phys}}}.	\\
	&= \frac{\bra{\Psi} \hat R^{-1}(\tau) \hat \Pi^{a}_{S} R(\tau) \hat \Pi^{b}_{SC}(\tau') \hat R^{-1}(\tau) \hat \Pi^{a}_{S} R(\tau) \ket\Psi_{\text{phys}}}{\bra\Psi \hat R^{-1}(\tau) \hat \Pi^{b}_{S} \hat R(\tau) \ket \Psi_{\text{phys}}} \\
	&= \frac{\bra{\psi(\tau)} \hat \Pi^{a}_{S} R(\tau) \hat \Pi^{b}_{SC}(\tau') \hat R^{-1}(\tau) \hat \Pi^{a}_{S} \ket{\psi(\tau)}}{\bra{\psi(\tau)} \hat \Pi^{a}_{S}  \ket{\psi(\tau)}}  \\
	&= \frac{\bra{\psi(\tau)} \hat \Pi^{a}_{S}R(\tau) \hat R^{-1}(\tau')\hat \Pi^{b}_{S}R(\tau') \hat R^{-1}(\tau)\hat \Pi^{a}_{S} \ket{\psi(\tau)}}{\bra{\psi(\tau)} \hat \Pi^{a}_{S}  \ket{\psi(\tau)}} \\
	&= \frac{\bra{\psi(\tau)} \hat \Pi^{a}_{S} \hat U(t-\tau') \hat \Pi^{b}_{S}U(\tau'-\tau) \hat \Pi^{a}_{S} \ket{\psi(\tau)}}{\bra{\psi(\tau)}\hat \Pi^{a}_{S}  \ket{\psi(\tau)}}
	\label{HSLresult}
\end{align} 
where we have used equation \cref{reduction} and $\hat R(\tau') \hat R^{-1}(\tau) = \hat U_S(\tau'-\tau)$. 
We have defined the "physical inner product"  as 
\begin{equation}
\braket{\Phi}{\Psi}_{phys} = \bra{Phi} \ketbra{s}{s}_C \otimes \mathbb{1}_S \ket{\Psi}.
\end{equation}
This inner product was introduced in \cite{Smith_Ahmadi_2019,Hoehn_2021} and shown to be independent of the time $s$.
\Cref{HSLresult} is indeed the correct two time conditional measurement probability as postulated by Born's rule. We now show that the \GLMshort  approach yields the same result: substituting the state in \cref{GLMstate} (with $\hat K_a = \hat \Pi^{a}_{S}$) into equation \cref{condGLM} yields 
\begin{align}
	P((b|\tau')|(a|\tau)) &= \frac{\big |\bra{\tau'}_C\tens \bra{b}_{M'}\tens\bra{a}_M)\hs\big |^2}{\big | {(\bra{\tau}_C\tens\bra{a}_M)\hs \big |}^2} \\
	&= \frac{\norm{\hat \Pi^{b}_{S} \hat U(\tau'-\tau) \hat \Pi^{a}_{S}\ket{\psi(\tau)}}^2}{\norm{\hat \Pi^{a}_{S}\ket{\psi(\tau)}}^2},
\end{align} 
identical to \cref{HSLresult}. It is straightforward to generalize this result to more than two subsequent measurements. The \GLMshort and \HSLshort approach hence result in the same physical predictions for perfect clocks, despite constructing a different physical picture!

\section{Calculations for the \GLM approach with non-ideal clocks}\label{appendix:calc_imperfect}
\subsection{The evolution equation for non-ideal clocks}\label{sec:tnlschrödinger}

In a similar manner to how the Schrödinger equation was derived in the ideal clocks case, we can also derive the Schrödinger equation in the non-ideal case. Let us remember that the Hamiltonian we consider in the non-ideal case is given by
\begin{equation}\label{eq:contraint_impefect}
    \hat H = \hat H_C + \hat H_S + N_{C} \int dt \ketbrat{t} \otimes \K t
\end{equation}
where $N_{C}$ is a normalization constant such that $\{N_C \ketbrat{t}\}_{t \in \mathbb{R}}$ forms a POVM.

As $\{\kett{t}\}_{t \in \mathbb{R}}$ is a over-complete basis, if 
\begin{equation}
     \brat{t} \hat H \ket{\Psi}_{CSA} = 0 \quad \forall t,
\end{equation} 
then this is equivalent to 
\begin{equation}
    \hat H \ket{\Psi}_{CSA} = 0.
\end{equation} 
Therefore, we consider the following problem, where for the sake of simplicity we absorbed the Hamiltonian of the system into the interaction term
\begin{equation}
     \brat{t}  \left( \hat H_C + N_{C}\int d\tau \ketbrat{\tau} \otimes \K\tau \right) \ket{\Psi}_{CSA} = 0.
\end{equation} 
As $\kett{t} = e^{- i \hat H_C t} \kett{0}$ is a coherent system of states it holds that $- i \frac{d}{dt} \brat{t} =\brat{t} \hat H_C$. Therefore, we find
\begin{equation}
      -i \frac{d}{dt}\brat{t} \ket{\Psi}_{CSA} + N_{C}\int d\tau \brakett{t}{\tau} \K\tau \brat{\tau}\ket{\Psi}_{CSA}  = 0.
\end{equation}
Defining $\brat{t} \ket{\Psi}_{CSA} = \ket{\psi(t)}_{SA}$ we find the following evolution equation:
\begin{equation}\label{eq:nl_evolution}
      -i \frac{d}{dt}\ket{\psi(t)}_{SA} + N_{C} \int d\tau \brakett{t}{\tau}   \K \tau \ket{\psi(\tau)}_{SA}  = 0.
\end{equation}

When considering solutions to this equation we have to be careful, as contrary to the ideal case, it is not clear that one can directly construct a solution to the constraint from them because in the case of non-ideal clocks the basis $\{\kett{t} \}_{t}$ is over-complete. 
The natural way to extend a solution $\ket{\psi(\tau)}_{SA}$ to the evolution equation \cref{eq:nl_evolution} on $SA$ to a solution for the constraint, would be to consider the state defined by
\begin{equation}\label{eq:psi_psi(t)}
    \ket{\Psi}_{CSA} = N_C \int d\tau \ \kett{\tau} \otimes \ket{\psi(\tau)}_{SA}.
\end{equation}
It is now not clear that the state $\ket{\Psi}_{CSA}$
satisfies the constraint, even though $\ket{\psi(\tau)}_{SA}$ satisfies the evolution equation \cref{eq:nl_evolution} as it is not clear that 
\begin{equation}
  \ket{\tilde{\psi}(t)}_{SA} = N_{C}\brat{t} \int d\tau \ \kett{\tau} \otimes \ket{\psi(\tau)}_{SA}.
\end{equation}
is again a solution to \cref{eq:nl_evolution}, even if $\ket{\psi(t)}_{SA}$ is.

In the case where $\ket{\psi(t)}_{SA} = \brat{t}\ket{\Psi}_{CSA}$, with
$\ket{\Psi}_{CSA}$ defined as in \cref{eq:psi_psi(t)}.
We find
\begin{align}
   \ket{\tilde{\psi}(t)}_{SA} &= N_{C}\brat{t}\int d\tau \ \kett{\tau} \otimes \ket{\psi(\tau)}_{SA} \\
   &= N_{C}\brat{t} \int d\tau \ \kett{\tau} \otimes \brat{\tau}\ket{\Psi}_{CSA} \\
   &= \brat{t} \ket{\Psi}_{CSA} =  \ket{\psi (t)}_{SA}.
\end{align}
For such a $\ket{\psi (t)}_{SA}$ we have that $\ket{\Psi}_{CSA}$, as defined in \cref{eq:psi_psi(t)}, fulfills the constraint \cref{eq:contraint_impefect}. 
As we shall see in the next section, the solution $\ket{\psi(t)}_{SA}$ does satisfy $\ket{\psi(t)}_{SA}=\bra{\phi_t}_C \ket{\Psi}_{CSA}$, with $\ket{\Psi}_{CSA}$ defined according to \cref{eq:psi_psi(t)}, thereby being a solution to \cref{eq:contraint_impefect}.

\subsection{Solving the evolution equation}
To solve \cref{eq:nl_evolution} we use a perturbative procedure. 
We then show that in the case where the approximation converges the solution, $\ket{\psi(t)}_{SA}$ fulfils the evolution equation and also that $\brat{t} \int d\tau \kett{\tau} \ket{\psi(\tau)}_{SA} = \ket{\psi(t)}_{SA}$.

Integrating the evolution equation leads us to the following equation
\begin{equation}\label{eq:int_ev_eq}
      \ket{\psi(t)}_{SA} = -i N_{C} \int_{-\infty}^{t} d\sigma \int d\tau \brakett{\sigma}{\tau}  \otimes \K \tau \ket{\psi(\tau)}_{SA} + \ket{\psi_0}_{SA}.
\end{equation}

To solve this equation, we make the approximation that the system does not change over time 
\begin{align}
    \ket{\psi(t)}^0_{SA}& := \ket{\psi_0}_{SA}
\end{align}
and then use the RHS of \cref{eq:int_ev_eq} to obtain a (hopefully) better approximation to the solution
\begin{align}
    \ket{\psi(t)}^1_{SA}&= \ket{\psi_0}_{SA} -i N_{C}\int_{-\infty}^{t} d\sigma \int d\tau \brakett{\sigma}{\tau}  \otimes \K\tau \ket{\psi_0}_{SA}.
\end{align}
To simplify this expression, we define
\begin{equation}
    F(t)= N_C \int_{-\infty}^{t} d\sigma \brakett{\sigma}{0}.
\end{equation}
Using the fact that 
\begin{equation}
    F(t-\tau)= N_C \int_{-\infty}^{t-\tau} d\sigma \brakett{\sigma}{0} =  N_C \int_{-\infty}^{t} d \sigma \brakett{\sigma-\tau}{0} = N_C \int_{-\infty}^{t} d \sigma \brakett{\sigma}{\tau}
\end{equation}
we find that 
\begin{align}
    \ket{\psi(t)}^1_{SA}&= \ket{\psi_0}_{SA} -i \int d\tau F(t-\tau)  \K\tau \ket{\psi_0}_{SA}.
\end{align}

We then use this new approximation as an Ansatz and repeat the procedure, to obtain approximations in higher orders of $\K t$
\begin{align}
\begin{split}
    \ket{\psi(t)}^2_{SA}&= \ket{\psi(t)}^1_{SA}+ (-i)^2 \int d \tau_1  \int d \tau_n F(t-\tau_1) F(\tau_1-\tau_2) \K {\tau_1} \K {\tau_2} \ket{\psi_0}_{SA} \\
    \dots \\
    \ket{\psi(t)}^n_{SA}&= \ket{\psi(t)}^{n-1}_{SA} \\ & \quad + (-i)^n \int d \tau_1 \dots  \int d \tau_n F(t-\tau_1) \dots F(\tau_{n-1}-\tau_n) \K {\tau_1} \dots \K {\tau_n }\ket{\psi_0}_{SA}
\end{split}
\end{align}

If this recursive procedure converges we obtain the following expression
\begin{equation}\label{eq:sol_evolution}
    \ket{\psi(t)}_{SA} = \sum_{n = 0}^{\infty} (-i)^n \int d \tau_1 \dots  \int d \tau_n F(t-\tau_1) \dots F(\tau_{n-1}-\tau_n) \K {\tau_1} \dots \K {\tau_n} \ket{\psi_0}_{SA}.
\end{equation}
In the case when the clock is ideal, i.e. $\brakett{t}{t'} = \delta(t-t')$, this reduces to the solution to the usual time-dependent Schrödinger equation, as $F(t) = N_C \int_{-\infty}^t d\sigma \brakett{\sigma}{0} = \Theta(t)$, the Heaviside theta function. 

Let us show that $\ket{\psi(t)}_{SA}$ is actually a solution to the evolution equation
\begin{align}
\begin{split}
    i \frac{d}{dt} \ket{\psi(t)}_{SA} &= \sum_{n = 1}^{\infty} (-i)^{(n-1)} \int d\tau_1 \dots d\tau_n N_C \brakett{t}{\tau_1} F(\tau_1-\tau_2) \dots F(\tau_{n-1}-\tau_{n}) \\ & \quad \K{\tau_{1}}\dots \K{\tau_{n}} \ket{\psi_0}_{SA}\\
   &= N_C \int d\tau \brakett{t}{\tau} \K{\tau} \sum_{n = 1}^{\infty} (-i)^{(n-1)} \\ &\quad \times \int d \tau_1  \dots d\tau_{n-1}  F(\tau-\tau_1) \dots F(\tau_{n-2}-\tau_{n-1}) \K{\tau_{1}}\dots \K{\tau_{n-1}} \ket{\psi_0}_{SA} \\
    &= N_{C} \int d\tau \brakett{t}{\tau} \K \tau \ket{\psi(\tau)}_{SA}.
\end{split}
\end{align}

We have now shown that the function $\ket{\psi(t)}_{SA}$ solves the evolution equation.
However, we have argued above that this does not yet imply that $\int dt \kett{t} \ket{\psi(t)}_{SA}$ solves the constraint,
as $\ket{\psi(t)}_{SA}$ does not necessarily have the form $\brat{t}\ket{\Psi}_{CSA}$.
For the type of Hamiltonian we consider here, we can show that this is indeed the case.
Let us define $\ket{\Psi}_{CSA} = N_C \int dt \kett{t} \ket{\psi(t)}_{CSA}$ and let us calculate $\brat{t}\ket{\Psi}_{CSA}$:
\begin{align}
\begin{split}
    \brat{t}\ket{\Psi}_{CSA} &= N_{C} \int d\tau \brakett{t}{\tau} \ket{\psi(\tau)}_{SA} \\
    &=  N_{C}\int d\tau \brakett{t}{\tau} \ket{\psi_0}_{SA} \\
    &\quad + \sum_{n = 1}^{\infty} (-i)^n N_{C} \int d\tau \int d \tau_1 \dots  \int d \tau_n \brakett{t}{\tau} F(\tau-\tau_1) \dots F(\tau_{n-1}-\tau_n)\\&\quad \times \K{ \tau_1 }\dots \K {\tau_n} \ket{\psi_0}_{SA}.
\end{split}
\end{align}
For the $n = 0$ term, using the expression of $\ket{\phi_\tau}$ in the Fourier basis, we find 
\begin{equation}
     N_{C} \int d\tau \brakett{t}{\tau} \ket{\psi_0}_{SA} = N_{C} \int d\tau \brakett{0}{\tau} \ket{\psi_0}_{SA}= \frac{1}{2 \pi}\int d \tau \int_{\sigma} d \epsilon \ e^{- i \tau \epsilon} \ket{\psi_0}_{SA} = \ket{\psi_0}_{SA}
\end{equation}
\\

\noindent 
For terms with $n > 0$, we consider the following expression, defining $f(t) = N_C \brakett{t}{0} = \frac{1}{2 \pi}\int_{\sigma} de \ e^{iet}$:
\begin{align}
    \begin{split}
        N_{C} \int d\tau \brakett{t}{\tau} F(\tau- \tau_1) &=   \int d\tau f(t-\tau) F(\tau- \tau_1) \\ &=  \int d\tau f(t-\tau_1-\tau) F(\tau) =  (f* F)(t-\tau_1).
    \end{split}
\end{align}
Using the relationship of convolution with integration, we find
\begin{equation}
    (f* F)(t-\tau_1) = \int_{-\infty}^{t-\tau_1} (f*f)(\tau) d\tau
\end{equation}
By definition of the function $f$, its Fourier transform is $
\mathcal{F}(f) = \chi_{\sigma}$, which implies that $\mathcal{F}(f*f) = \mathcal{F}(f)^2 = \mathcal{F}(f)$ and thus $(f*f)(\tau) = f(\tau)$.
Together this implies that
\begin{equation}
   N_{C} \int d\tau \brakett{t}{\tau} F(\tau- \tau_1) = F(t-\tau_1).
\end{equation}
Plugging this back into our expression for $\brat{t}\ket{\Psi}_{CSA}$ we find
\begin{align}
    \brat{t}\ket{\Psi}_{CSA} = \ket{\psi(t)}_{SA},
\end{align}
which proves that the solution we found satisfies the constraint. 

\subsection{The evolution equation and its solution for periodic clocks}
If we want to consider a finite-dimensional clock we need to consider a periodic time, 
where for the clock there exists a time $T$ such that $\kett{T} = \kett{0}$ \cite{Hoehn_2021}. From integrating evolution equation
\begin{equation}\label{eq:periodic_constraint}
      \ket{\psi(t)}_{SA} =(-i) N_{C} \int_{0}^{t} d\sigma \int_0^T d\tau \brakett{\sigma}{\tau}  \K \tau \ket{\psi(\tau)}_{SA} + \ket{\psi_0}_{SA}. 
\end{equation}
Applying the same technique as for the non-periodic case we find the following evolution if the recursive procedure converges
\begin{align}
\begin{split}
     \ket{\psi(t)}_{SA} &= \sum_{n = 0}^{\infty} (-i)^n \int_{0}^{t} d\sigma_1 \int_{0}^T d \tau_1 \dots \int_{0}^{\tau_{n-1}} d\sigma_n \int_0^T d \tau_n \brakett{\sigma_1}{\tau_1} \dots \brakett{\sigma_n}{\tau_n} \\& \quad \times  \K \tau_1 \dots K \tau_n \ket{\psi_0}_{SA}
\end{split}
\end{align}
For simplicity of notation, we define 
\begin{equation}
    F_{\tau}(t) := N_C \int_{0}^{t} d\sigma \brakett{\sigma}{\tau}.
\end{equation}
With this definition, we can write the solution in a more compact form
\begin{equation}
    \ket{\psi(t)} = \sum_{n = 0}^{\infty} (-i)^n  \int_{0}^T d \tau_1 \dots \int_0^T d \tau_n F_{\tau_1}(t) \dots F_{\tau_n}(\tau_{n-1}) \K{\tau_1} \dots \K{\tau_n}\ket{\psi_0}_{SA}.
\end{equation}
As in the non-periodic case, we find a candidate for the solution to \cref{eq:periodic_constraint} by appending the clock and integrating over time
\begin{equation}\label{definitionperiodic}
    \ket{\Psi}_{CSA} = N_{C} \int d\tau \kett{\tau} \otimes\ket{\psi(\tau)}_{SA}.
\end{equation}
Note that $\ket{\Psi}_{CSA}$ does not necessarily need to fulfil the constraint.
In this case, we will see that $\brat{t}\ket{\Psi}_{CSA}$ is a solution to the evolution equation and, thus, $\ket{\Psi}_{CSA}$ satisfies the constraint if the solution $\ket{\psi(t)}_{SA}$ is periodic and $\brat{0}\ket{\Psi}_{CSA} = \ket{\psi(0)}_{SA}$. 
Let us calculate $\brat{t}\ket{\Psi}_{CSA}$:
\begin{align}
    \brat{t}\ket{\Psi}_{CSA} &= N_{C} \int_0^T d\tau \brakett{t}{\tau} \ket{\psi(\tau)}_{SA}.
\end{align}
We derive this expression by $t$ and find
\begin{equation}\label{periodicevequation}
   \frac{d}{dt} \brat{t}\ket{\Psi}_{CSA} = N_{C} \int_0^T d\tau \brakett{t}{\tau} \frac{d}{d\tau}\ket{\psi(\tau)}_{SA} - N_{C} \brakett{t}{0}\left(\ket{\psi(T)}_{SA}-\ket{\psi(0)}_{SA}\right).
\end{equation}
Using that $\ket{\psi(t)}_{SA}$ is a solution to the evolution equation we find
\begin{equation}
   \frac{d}{dt} \brat{t}\ket{\Psi}_{CSA} = \frac{d}{dt}\ket{\psi(\tau)}_{SA} - N_{C}\brakett{t}{0}\left(\ket{\psi(T)}_{SA}-\ket{\psi(0)}_{SA}\right).
\end{equation}
If we assume that the solution $\ket{\psi(t)}_{SA}$ is periodic, this implies
\begin{equation}
   \frac{d}{dt} \brat{t}\ket{\Psi}_{CSA} = \frac{d}{dt}\ket{\psi(\tau)}_{SA}.
\end{equation}
Integrating both sides of the equation we find
\begin{equation}
    \brat{t}\ket{\Psi}_{CSA} = \ket{\psi(t)}_{SA} +\ket{\psi(0)}_{SA} - N_{C} \int_0^T d\tau \brakett{0}{\tau}  \ket{\psi(\tau)}_{SA}.
\end{equation}
Therefore $\brat{t}\ket{\Psi}_{CSA}$ satisfies the constraint if
\begin{equation}
    \ket{\psi(0)}_{SA} - N_{C} \int_0^T d\tau \brakett{0}{\tau} \ket{\psi(\tau)}_{SA} = 0.
\end{equation}

\section{From purified measurements to twirled observables}\label{appendix:glm_to_hsl}
\noindent The probability of obtaining measurement outcome $a$ at time $t$ for a measurement event at time $\tau$ in the \GLMshort approach can be written as  
\begin{equation}\label{app_prob_glm}
P(a|t, \tau) = \frac{ \mathrm{Tr}( \ketbra{\phi_t}{\phi_t}_C \otimes \mathbb{1}_S \otimes \ketbra{a}{a}_A  \ketbra{\Psi}_{CSA})}{\mathrm{Tr}( \ketbra{\phi_t}{\phi_{t}}_C \otimes \mathbb{1}_{SA} \ketbra{\Psi}_{CSA})},
\end{equation}
where $\ket{\Psi}$ is a solution to the constraint 
\begin{equation}\label{const_imperfect_glm}
    \hat H_{CSA} \ket{\Phi}_{CSA} = \left(\hat H_C + \hat H_S + N_{C} \int dt \ketbrat{t} \otimes \K_{SA}{t} \right) \ket{\Phi}_{CSA} = 0,
\end{equation}
with a measurement interaction at time $\tau$. 

We can express the solution as $\ket{\Psi}_{CSA}=\piGlm\ket{\psi_0}_{CSA}$, where $\piGlm$ is the "projector" onto the space of solutions to the constraint in \cref{const_imperfect_glm}
\begin{equation}
   \hat \Pi_{PM} = \int_{\mathbb{R}} dt e^{-i t \hat H_{CSA}},
\end{equation}
where $\hat{H}$ is given by \cref{const_imperfect_glm}.
The state $\ket{\psi_0}_{CSA}$ can be interpreted as an initial condition and is an arbitrary element of the kinematical Hilbert space. We extend the \HSLshort approach to obtain the ancillary system $A$, and assume that $A$ contributes trivially to the \HSLshort constraint of \cref{PerfectConstraint}. Then, a solution $\ket{\eta}_{CSA}$ to the \HSLshort constraint now reads
\begin{equation}
(\hat H_C + \hat H_S)\ket{\eta}_{CSA} = 0,
\end{equation}
with identities on $A$ left implicit. 
We choose $\ket{\eta}_{CSA}$ such that $\ket{\psi_0}_{CSA} = \ketbra{\phi_0}{\phi_0}_C \ket{\eta}_{CSA}$. 
Then, using the cyclicity of the trace and the fact that $\ketbra{\eta}{\eta}_{CSA}$ is left unchanged under incoherent G-twirling $\mathcal{T}$ (with respect to the \HSLshort time translation), we can rewrite \cref{app_prob_glm} as
\begin{equation}\label{translation}
P(a|t,\tau) = \frac{ \mathrm{Tr}\left(\, \mathcal{T} [ \ketbra{\phi_0}{\phi_0}_C \piGlm^{\dagger} \ketbra{\phi_t}{\phi_t}_C \otimes \ketbra{a}{a}_A \piGlm \ketbra{\phi_0}{\phi_0}_C ] \ketbra{\eta}{\eta}_{CSA}\right)}{\mathrm{Tr} \left(\, \mathcal{T} [ \ketbra{\phi_0}{\phi_0}_C \piGlm^{\dagger} \ketbra{\phi_t}{\phi_t}_C \piGlm \ketbra{\phi_0}{\phi_0}_C ] \ketbra{\eta}{\eta}_{CSA}\right)},
\end{equation}
where we used that $\mathcal{T}$ is self-adjoint and have left identity operators implicit for simplicity. 

The operators in the numerator and denominator of \cref{translation} are explicitly invariant under the action generated by the \HSLshort constraint. 
Moreover, the state $\ket{\eta}$ satisfies the \HSLshort constraint.
Therefore, \cref{translation} is a conditional probability in the spirit of the \HSLshort approach. 
It remains to be seen, however, whether this probability can be written exactly in terms of the "physical inner product" (see \cref{appendix:glm=hls}) of the \HSLshort approach. 

In general, the operators in \cref{translation} are not of the form 
\begin{equation}\label{HSLop}
\hat F_{SC}(\tau) =  \int_{\mathbb R} dt \, \ e^{- i t\hat H_C} \, \ketbra{\tau}{\tau}_C \, e^{ i t\hat H_C}\tens e^{- i t\hat H_S}\, \hat F_S \, e^{ i t\hat H_S},
\end{equation} 
which represents time-local measurements. 
Unless we are in the limit where clocks can be approximated as perfect, these operators represent measurements that are delocalised in time in the \HSLshort approach. 

\section{Discrete time model}\label{appendix:discrete_time}
In the case the spectrum of the clock is the interval $[-E,E]$, even though, in general, clock states are not orthogonal, there still exists times $\{t_k\}_{k}$, where the clock states are mutually orthogonal $\braket{t_k}{t_{k'}} = \delta_{k,k'}$.\footnote{This approach generalizes to other clock Hamiltonians that have times at which the clock states are mutually orthogonal.} Using the explicit form of the spectrum, we find
\begin{equation}
   \braket{t_k}{t_{k'}} = \frac{1}{2 \pi} \int_{-E}^{E} d e \ \e^{i (t_k - t_{k'}) e} = \frac{E}{\pi} \sinc({E (t_k-t_{k'})})
\end{equation}
with $\sinc(x) := \frac{\sin(x)}{x}$. So the clock states are orthogonal $\braket{t_k}{t_{k'}} = 0$, if $t_k-t_{k'} = \frac{N \pi}{E}$ with $N \in \mathbb{Z}$. Therefore, we can choose the following orthonormal clock basis 
\begin{equation}
    \left\{\ket{t_k} \middle| \ t_{k} = \frac{k \pi}{E} \ k \in \mathbb{Z}\right\}.
\end{equation}

The unitary 
\begin{equation}
    \hat U = \sum_{k} \ketbra{t_{k+1}}{t_k} \otimes \hat U^{(k)}_{SA}
\end{equation}
can be interpreted as the unitary evolving the system and the clock for one time step, such that the system undergoes the evolution $U_k$ when the clock shows time $t_k$. By choosing the unitary $U^{(k^*)}$ for some integer $k^*$ to be an interaction for a von Neumann unitary, then we can interpret this as the system being measured at time $t_{k^*}$. 

In this discrete model, we no longer have a Hamiltonian, but the evolution of the system for a single time step. The constraint we considered in the continuous case, is equivalent to the total system not evolving in time:
\begin{equation}
    \e^{i \hat H_{CSA} t} \ket{\Psi}_{CSA} = \ket{\Psi}_{CSA} \quad \forall \ t.
\end{equation}
So for the discrete case, we impose the following constraint:
\begin{equation}
    \hat U^{n} \ket{\Psi}_{CSA} = \ket{\Psi}_{CSA} \quad \forall n \in \mathbb{Z}.
\end{equation}
This is equivalent to 
\begin{equation}
    \hat U \ket{\Psi}_{CSA} = \ket{\Psi}_{CSA}.
\end{equation}
Expanding the state $\ket{\Psi}_{CSA}$ in the discrete clock basis we find the condition
\begin{equation}
    \sum_{k} \ket{t_{k+1}}_C \otimes \hat U^{(k)}_{SA}\ket{\psi(t_k)}_{SA} =  \sum_{k} \ket{t_k}_C \otimes \ket{\psi (t_k)}_{SA},
\end{equation}
from which we find that the states $\ket{\psi(t_k)}_{SA}$ need to fulfil the relation
\begin{equation}
    \ket{\psi(t_{k+1})}_{SA} = \hat U^{(k)}_{SA} \ket{\psi(t_k)}_{SA}.
\end{equation}
Applying this relation recursively we find
\begin{equation}
    \ket{\psi(t_{k+1})}_{SA} = \prod_{j = 0}^k \hat U^{(j)}_{SA} \ket{\psi(t_{0}}_{SA}.
\end{equation}
We then see that, in this model, the evolution of the system relative to the clock is unitary as long as we consider time steps of size $\frac{\pi}{E}$.

Moreover, for discrete times, it follows immediately that the \GLMshort approach constructed above and the discrete version of the \HSLshort approach, defined in the obvious way, produce equivalent predictions.

\end{document}